\documentclass[a4paper,11pt]{article}
\pdfoutput=1 
\usepackage{jheppub} 
\usepackage{amsmath,bbm,graphicx,bm,slashed,multirow,color}
\usepackage[T1]{fontenc}

\DeclareMathOperator\Tr{Tr}
\DeclareMathOperator\tr{tr}

\newcommand{\1}{1\hspace{-2.7pt}\textrm{l}}
\newcommand{\del}{\partial}
\newcommand{\SU}{\text{SU}}
\newcommand{\U}{\text{U}}

\newcommand{\ee}{\,\mathrm{e}}
\newcommand{\RF}{R^F_k}
\newcommand{\RB}{R^B_k}
\newcommand{\SO}{\text{SO}}

\newcommand{\DD}{\slashed{D}}

\newcommand{\pp}{\hat{p}}
\newcommand{\qq}{\textbf{q}}
\newcommand{\FF}{\textbf{F}}

\newcommand{\zetax}{\hat{m}^2_{\sigma}}
\newcommand{\zetay}{\hat{m}^2_{\pi}}

\newcommand{\zetasp}{\hat{m}_{\sigma,\pi}}

\renewcommand{\O}{\text{O}}
\renewcommand{\epsilon}{\varepsilon}
\renewcommand{\bar}[1]{\overline{#1}} 

\title{Chiral dynamics in a magnetic field from the functional renormalization group}

\author[a]{Kazuhiko Kamikado}
\author[b]{and Takuya Kanazawa}

\affiliation[a]{Theoretical Research Division, Nishina Center, RIKEN, Saitama 351-0198, Japan}
\affiliation[b]{Quantum Hadron Physics Laboratory, RIKEN, Saitama 351-0198, Japan}

\emailAdd{kazuhiko.kamikado@riken.jp}
\emailAdd{takuya.kanazawa@riken.jp}

\abstract{ We investigate the quark-meson model in a magnetic field
using the functional renormalization group equation beyond the
local-potential approximation.  Our truncation of the effective action
involves anisotropic wave function renormalization for mesons, which
allows us to investigate how the magnetic field distorts the propagation
of neutral mesons. Solving the flow equation numerically, we find that
the transverse velocity of mesons decreases with the magnetic field at
all temperatures, which is most prominent at zero temperature. The meson
screening masses and the pion decay constants are also computed. The
constituent quark mass is found to increase with magnetic field at all
temperatures, resulting in the crossover temperature that increases
monotonically with the magnetic field. This tendency is consistent with
most model calculations but not with the lattice simulation performed at
the physical point. Our work suggests that the strong anisotropy of
meson propagation may not be the fundamental origin of the inverse
magnetic catalysis.  }

\begin{document}
\maketitle
\flushbottom

 \section{Introduction}
 Understanding strongly coupled dynamics of Quantum Chromodynamics (QCD)
 from first principles is one of the most important challenges in modern
 theoretical physics. Chiral symmetry breaking and quark confinement are
 two hallmarks of the nonperturbative QCD vacuum. Moreover QCD exhibits
 novel phenomena under extreme conditions, such as color deconfinement
 at high temperature and color superconductivity at high baryon chemical
 potential. These areas are actively investigated in relation to the
 physics of compact stars, heavy ion collisions, and early Universe; see
 \cite{Fukushima:2010bq} for a review.

 Recently QCD in an external magnetic field has attracted considerable
 attention. The magnetic field is not only interesting as a theoretical
 probe to the dynamics of QCD, but also important in cosmology and
 astrophysics. A class of neutron stars called magnetars has a
 strong surface magnetic field of order $10^{10}$ T \cite{Duncan:1992hi}
 while the primordial magnetic field in early Universe is estimated to
 be even as large as $\sim 10^{19}$ T \cite{Grasso:2000wj}. In
 non-central heavy ion collisions at RHIC and LHC, a magnetic field of
 strength $\sim 10^{15}$ T perpendicular to the reaction plane could be
 produced and can have impact on the thermodynamics of the quark-gluon
 plasma \cite{Skokov:2009qp}.
 
 The effect of magnetic field has been vigorously investigated in chiral
 effective models
 \cite{Suganuma:1990nn,Klimenko:1990rh,Klimenko:1991he,Gusynin:1994re,Gusynin:1994va,
 Gusynin:1994xp,Gusynin:1995nb,Ebert:2003yk,Inagaki:2003yi,Fraga:2008qn,Menezes:2008qt,
 Menezes:2009uc,Boomsma:2009yk,Fukushima:2010fe,Mizher:2010zb,Fayazbakhsh:2010bh,
 Gatto:2010pt,Chatterjee:2011ry,Frasca:2011zn,Rabhi:2011mj,Kashiwa:2011js,Andersen:2011ip,
 Skokov:2011ib,Scherer:2012nn,Fukushima:2012xw,Andersen:2012bq,Fayazbakhsh:2012vr,Fayazbakhsh:2013cha,
 Ferrari:2012yw,Preis:2012fh,Andersen:2013swa,Ferrer:2013noa} (see
 \cite{Gatto:2012sp,Shovkovy:2012zn} for reviews).  It was found that
 the magnetic field acts as a catalyst of chiral symmetry breaking, an
 effect called \emph{magnetic catalysis}. This model-independent
 phenomenon is explained through dimensional reduction ($3+1 \to 1+1$)
 in the quark pairing dynamics in a magnetic field
 \cite{Gusynin:1994xp,Gusynin:1995nb}.

 The dynamics of QCD in a magnetic field has also been studied in
 lattice simulations
 \cite{Buividovich:2008wf,D'Elia:2010nq,D'Elia:2011zu,Braguta:2011hq,Bali:2011qj,Ilgenfritz:2012fw,
 Luschevskaya:2012xd,Bali:2012zg,Bali:2012jv,Bali:2013esa,Bonati:2013lca,Levkova:2013qda,
 Ilgenfritz:2013ara,Bonati:2013vba}, see \cite{D'Elia:2012tr} for a
 review. At a relatively large quark mass, the chiral condensate and the
 chiral restoration temperature were found to increase with the magnetic
 field in accordance with the magnetic catalysis scenario,%
 \footnote{However, \emph{inverse} magnetic catalysis with large quark mass was reported quite recently \cite{Bornyakov:2013eya}.} whereas
 simulations at the physical quark masses \cite{Bali:2011qj,Bali:2012zg}
 show that the effect of a magnetic field is non-monotonic: the chiral
 condensate increases at low temperature, but \emph{decreases} at high
 temperature, resulting in a lower pseudo-critical temperature in a
 stronger magnetic field. The origin of this \emph{inverse magnetic
 catalysis} (or \emph{magnetic inhibition}) is not fully understood yet.

 Possible explanations for the inverse magnetic catalysis have been
 suggested by several groups
 \cite{Fukushima:2012kc,Kojo:2012js,Bruckmann:2013oba,Chao:2013qpa}. Among others,
 Fukushima and Hidaka \cite{Fukushima:2012kc} noted that the dimensional
 reduction of neutral pion could be a source of disorder that weakens
 chiral symmetry breaking. The idea is rooted in the observation that
 the neutral pion `feels' the magnetic field through its internal quark
 and anti-quark, and consequently the pion can move in directions
 transverse to the magnetic field with little energy cost
 \cite{Gusynin:1994xp,Gusynin:1995nb,
 Fayazbakhsh:2012vr,Fayazbakhsh:2013cha}. However the analysis of
 \cite{Fukushima:2012kc} was limited to zero temperature, and the impact
 of anisotropic fluctuations of neutral pion on the finite-temperature
 dynamics of QCD has not been quantitatively investigated.
 
 In this work, we apply the functional renormalization group (FRG)
 \cite{Wetterich:1992yh} to the quark-meson model to study chiral
 symmetry breaking and its restoration at finite temperature under a
 magnetic field. FRG is a powerful nonperturbative method to go beyond
 the mean-field approximation by fully taking thermal and quantum
 fluctuations into account. The basic idea of FRG is to start from a
 microscopic action at the UV scale $k=\Lambda$, and keep track of the
 flow of the scale-dependent effective action while integrating out
 degrees of freedom with intermediate momenta successively; finally at
 $k=0$ the full quantum effective action is obtained. See
 \cite{Berges:2000ew,Pawlowski:2005xe,Delamotte:2007pf,Braun:2011pp} for
 reviews. While FRG has already been applied to chiral models in a
 magnetic field
 \cite{Skokov:2011ib,Scherer:2012nn,Fukushima:2012xw,Andersen:2012bq,Andersen:2013swa},
 so far no attempt has been made to go beyond the leading order of the
 derivative expansion, known as the \emph{local-potential approximation}
 (LPA) in which the meson fluctuations are included but the
 scale-dependent flow of the kinetic term is entirely neglected. In this
 work, we proceed to the next order of the derivative expansion by
 including the wave function renormalization.  This enables us to
 investigate the \emph{strongly anisotropic} meson fluctuations for the
 first time. We will show that the pion decay constant and the meson
 screening masses become direction dependent, due to the breaking of the
 rotational symmetry by a magnetic field, and that the pion's
 \emph{transverse velocity} (i.e. the velocity in the direction
 perpendicular to the magnetic field) decreases significantly under a
 strong magnetic field.  \footnote{This is similar to the effect of the 
 heatbath in finite-temperature QCD where the temporal decay constant
 differs from the spatial decay constant and the pion velocity is less
 than the speed of light \cite{Pisarski:1996mt,Son:2001ff,Son:2002ci}.}
 To be specific, we will compute following quantities as functions of
 temperature and magnetic field strength:
 \begin{itemize}
  \setlength{\itemsep}{-1.5mm}
  \item Constituent quark mass $(M_q)$
  \item Transverse meson screening masses $(m_{\pi,\sigma}^\perp)$
  \item Longitudinal meson screening masses $(m_{\pi,\sigma}^\parallel)$ 
  \item Transverse pion decay constant $(f_\pi^\perp)$
  \item Longitudinal pion decay constant $(f_\pi^\parallel)$
  \item Wave function renormalization factors for mesons ($Z^\perp$, $Z^\parallel$)
  \item Transverse velocity of mesons $(v^2_\perp \equiv Z^\perp/Z^\parallel)$
  \item Chiral restoration temperature $(T_{\rm pc})$
 \end{itemize}
 Our model calculations for the anisotropic screening masses and the
 transverse velocity of pions offer predictions that can be tested in
 future lattice simulations. As for the pseudo-critical temperature,
 contrary to the expectation from \cite{Fukushima:2012kc}, we did not
 observe agreement with lattice data: $T_{\rm pc}$ \emph{increases}
 monotonically with the magnetic field as in other model calculations,
 despite the fact that our present calculation incorporates
 significantly more meson fluctuations than other calculations. While
 our truncation of the effective action is still far from being complete
 and can be extended further, the discrepancy with lattice data could be
 taken as evidence that gluonic degrees of freedom which are ignored in
 chiral models actually play a vital role in the phenomenon of inverse
 magnetic catalysis.
 
 This paper is organized as follows.  In section \ref{sc:frg_qm} we
 introduce the quark-meson model and describe the formulation of FRG. We
 specify our truncation of the effective action and introduce regulators
 that are devised for analysis in a magnetic field.  Then we give full
 expressions for the flow equations (omitting the details of derivation)
 and discuss the setup to solve them numerically. In section
 \ref{sc:num_res} we show plots of physical observables obtained with a
 numerical method, discuss their characteristics, and compare with the
 mean-field treatment and LPA. We will also comment on agreement and
 discrepancy with the available lattice data. Section \ref{sc:concl} is
 devoted to conclusion. The analytical derivation of all the flow
 equations is presented in full details in appendices
 \ref{sc:der_flow_Uk}, \ref{sc:der_flow_Zk}, and \ref{sc:taylorU}.

 \section{Functional renormalization group for the quark-meson model}
 \label{sc:frg_qm}

 In this section we describe the setup of FRG for the quark-meson model
 in a magnetic field. In general, FRG requires specification of the
 following 4 ingredients: (1) the flow equation, (2) regulator
 functions, (3) truncation of the effective action, and (4) initial
 conditions for the flow. We will describe (1)--(3) in this section and
 (4) in section \ref{sc:parafix}.

  \subsection{General structure of the flow and regulators} 
  The functional renormalization group equation (called the Wetterich
  equation) reads
  \begin{align}
   \label{eq:FRG-}
   \del_k \Gamma_k = 
   {\frac{1}{2}\Tr\left[ \frac{1}{\Gamma^{(2,0)}_k+\RB} \del_k \RB \right]}
   - {\Tr\left[ \frac{1}{\Gamma^{(0,2)}_k+\RF} \del_k \RF \right]}\,, 
  \end{align}
  which describes the evolution of the scale-dependent effective action
  $\Gamma_k$ from the initial UV scale ($k=\Lambda$) to the IR limit
  ($k=0$).  $\Gamma_{k=\Lambda}$ is taken to be equal to the classical
  action and $\Gamma_{k=0}$ is the full quantum effective action
  incorporating the effects of all fluctuations.  Here $\RB$ and $\RF$
  are cutoff functions (regulators) for bosons and fermions, while
  $\Gamma_k^{(2,0)}$ and $\Gamma_k^{(0,2)}$ represent the second
  functional derivative of $\Gamma_k$ with respect to boson fields and
  fermion fields, respectively.  $\Tr$ is a trace in the functional
  space. Further details on FRG can be found in reviews
  \cite{Berges:2000ew,Pawlowski:2005xe,Delamotte:2007pf,Braun:2011pp}.
    
  Although \eqref{eq:FRG-} has a simple one-loop structure, it must be
  distinguished from the perturbative one-loop approximation: actually
  \eqref{eq:FRG-} incorporates effects of arbitrarily high order
  diagrams in the perturbative expansion through the full
  field-dependent propagator \mbox{$(\Gamma^{(2)}_k+R_k)^{-1}$}.
        
  The flow of $\Gamma_k$ from UV to IR is controlled by the cutoff
  functions $R_k^{B,F}(p)$.  The latter must satisfy (i) $\displaystyle
  \lim_{k\to\infty}R_k(p)=\infty$, (ii) $\displaystyle \lim_{k\to
  0}R_k(p)=0$, and (iii) $\displaystyle \lim_{p\to 0}R_k(p)>0$
  \cite{Berges:2000ew}.  In this work we use the following
  \emph{anisotropic} regulators
  \begin{align}
   \label{eq:RB_3d}
   \RB(p) & = (k^2-p_3^2) Z_k^\parallel \,\theta(k^2-p_3^2) \,,
   \\
   \RF(p) & 
   = - i \slashed{p}_3 r_k(p_3)\qquad \text{with }\quad 
   r_k(p_3) \equiv \left(\frac{k}{|p_3|}-1\right) \theta(k^2-p_3^2) \,,
   \label{eq:RF_3}
  \end{align}
  for bosons and fermions ($\slashed{p}_3=p_3\gamma_3$),
  respectively. Here $Z_k^\parallel$ is a wave function renormalization
  factor for mesons (cf.~section \ref{sc:sdea}).  These regulators
  comply with the conditions (i)--(iii) above. Actually they are nothing
  but Litim's optimized regulator but now restricted to the $p_3$
  direction.  On one hand, these (somewhat unusual) regulators that
  break rotational symmetry are quite convenient because of a simple
  form of the scale-dependent fermion propagator in a magnetic field, as
  will be demonstrated later.  On the other hand, they render the flow
  equation UV-divergent as they do not suppress momenta $p_1$ and $p_2$
  at all.  We will return to this problem later. Associated with this,
  we remark that the scale-dependent action $\Gamma_k$ no longer admits
  a naive interpretation as a Wilsonian coarse-grained effective action
  at scale $k$, because the above regulators do not suppress modes with
  momenta $p_{1,2}^2\lesssim k^2$.  However, we hasten to add that those
  regulator functions work \emph{perfectly well} as a machinery to
  interpolate between the classical action and the full quantum
  effective action.

  \subsection{Scale-dependent effective action} 
  \label{sc:sdea}
  
  Next, let us define the model we use and specify our truncation of the
  running effective action.  If we consider realistic QCD with two
  flavors of charge $+2e/3$ and $-e/3$, the chiral symmetry
  $\SU(2)_R\times \SU(2)_L \cong \O(4)$ would be \emph{explicitly}
  broken even in the chiral limit and consequently the flow equation
  becomes highly complicated: the scale-dependent effective potential
  would no longer be a function of the single $O(4)$-symmetric variable
  $\sigma^2+\vec\pi^2$, \footnote{This point seems to have been
  neglected in earlier works
  \cite{Skokov:2011ib,Andersen:2012bq,Andersen:2013swa}.} and also the
  wave function renormalization factors for $\pi^{\pm}$ and $\pi^0$ will
  be different in general.

  To avoid these complications and focus on the mechanism proposed by
  Fukushima and Hidaka \cite{Fukushima:2012kc}, we will limit ourselves
  to the quark-meson model \cite{Jungnickel:1995fp,Berges:1997eu} with
  \emph{one flavor} of a fermion with charge $e$ and color $N_c$\,. (We
  ignore the axial anomaly.)  In this model, the pion ($\pi$) is
  neutral. Since $\pi^{\pm}$ in real $N_f=2$ QCD decouple from the
  low-energy dynamics in a strong magnetic field and only the neutral
  pion $\pi^0$ remains light, it essentially reduces to the model
  considered here.

  While the original Wetterich equation \eqref{eq:FRG-} formulated in
  the infinite-dimensional functional space is \emph{exact}, in practice
  we need to find a proper truncation of $\Gamma_k$ to make explicit
  computations feasible.  A variety of truncation schemes have been
  discussed in the literature.  Among others, the leading order of the
  derivative expansion, called the local-potential approximation (LPA),
  is frequently used due to its technical simplicity and was also
  employed in \cite{Skokov:2011ib,Andersen:2012bq,Andersen:2013swa}. In
  LPA the effective potential flows with $k$ while the field
  renormalization is neglected altogether, resulting in identically
  vanishing anomalous dimension of fields.  In this work, we go beyond
  LPA by employing the following truncation of the running effective
  action:
  \begin{align}
   \label{eq:TB}
   \Gamma_k [\psi, \sigma, \pi] = & \int_0^\beta \! dx_4 \int d^3x ~
   \Bigg \{ 
   \sum_{a=1}^{N_c} \bar\psi_a [ \DD + g (\sigma + i\gamma_5 \pi) ] \psi_a 
   + U_k(\rho) - h\sigma
   \notag
   \\
   & \qquad 
   + \frac{Z_k^\perp}{2} \sum_{i=1,2} [(\del_i \sigma)^2 + (\del_i \pi)^2] 
   + \frac{Z_k^\parallel}{2} \sum_{i=3,4} [(\del_i \sigma)^2 + (\del_i \pi)^2]  \Bigg\}\,,
  \end{align}
  with $\beta=1/T$ and $\rho\equiv \frac{1}{2}(\sigma^2+\pi^2)$.  The
  Dirac operator reads
  \begin{align}
   \DD = \gamma_\mu D_\mu\,, \qquad D_\mu = \del_\mu -ieA_\mu\,,\qquad
   \bm{A}=(0,Bx_1,0)\,, \quad \text{and} \quad  A_4=0\,.
  \end{align}
  One can verify that the action possesses $\U(1)\times \U(1)$ chiral
  symmetry when $h=0$. The parameter $h$ that enters as a symmetry
  breaking field parametrizes the effect of current quark mass. Below we
  assume $h>0$. In \eqref{eq:TB} we introduced the wave function
  renormalization factors $Z_k^\perp$ and $Z_k^\parallel$. Setting
  $Z_k^\perp=Z_k^\parallel=1$ brings us back to LPA. Here we let these
  variables depend on $k$. It is important that $Z_k^\perp$ for
  directions perpendicular to the magnetic field, and $Z_k^\parallel$
  for directions parallel to the magnetic field, are treated
  independently. This setup is well-motivated in view of the anisotropy
  induced by a magnetic field and is actually essential to test the
  scenario by Fukushima and Hidaka \cite{Fukushima:2012kc}.

  Several caveats are in order. Firstly, we neglect the wave function
  renormalization of fermions and the derivative term of $\rho$ (i.e.,
  $(\del_\mu \rho)^2$), as well as all bosonic terms that are consistent
  with symmetries and include more than two derivatives.  We also ignore
  the $k$-dependence of $g$ because the flow of $g$ is not expected to
  affect final results significantly (see e.g.,
  \cite{Jungnickel:1995fp}).  In principle all these corrections can be
  incorporated into the present approach in a straightforward
  manner,\footnote{See however \cite{Morris:2005ck,Morris:1999ba} for a
  subtlety in the higher-order derivative expansion based on a
  non-smooth regulator, such as Litim's optimized regulator.  This issue
  does not arise at the order of expansion considered in this paper.}
  but it is beyond the scope of this work. Secondly, for technical
  simplicity, we use a common variable, $Z_k^\parallel$, for both the
  wave function renormalization factor in $x_4$-direction and that in
  $x_3$-direction. We assume the error due to this approximation is
  small (see \cite{Braun:2009si} for a discussion on a related issue at
  finite temperature).

  \subsection{Flow equations for the quark-meson model}
  With \eqref{eq:FRG-}, \eqref{eq:RB_3d}, \eqref{eq:RF_3} and
  \eqref{eq:TB}, we are now ready to derive the flow equations for
  $U_k(\rho)$, $Z_k^\perp$ and $Z_k^\parallel$ explicitly. Since their
  analytical derivation is rather lengthy and involved, we shall
  relegate it to the appendices \ref{sc:der_flow_Uk} and
  \ref{sc:der_flow_Zk}.  Here we only quote the main formulas:
  \begin{align}
   &\del_k U_k(\rho) 
    = k^2 \left( 1 + \frac{k}{3}\frac{\del_kZ_k^\parallel}{Z_k^\parallel} \right) 
   {\int \,}'\frac{d^2p_\perp}{(2\pi)^3}\left(
   \frac{1}{E_\pi(\rho)}\coth\frac{E_\pi(\rho)}{2T} + 
   \frac{1}{E_\sigma(\rho)}\coth\frac{E_\sigma(\rho)}{2T}
   \right)
   \notag
   \\
   & \qquad \qquad \quad - \frac{1}{2\pi^2}N_ck^2  |eB|\ {\sum_{n=0}^{\infty}}'
   \frac{\alpha_n}{E_n(\rho)}\tanh \frac{E_n(\rho)}{2T} 
   \,,
   \label{eq:fin_dU}
   \\
   &{\del_k Z_k^\perp} 
    = \scalebox{0.95}{$\displaystyle 
   -  \frac{k^2}{\pi^2} \frac{\bar\rho_k\big[U''_k(\bar\rho_k)\big]^2}
   {(Z_k^\parallel)^2}
   \left(1 + \frac{k}{3} \frac{\del_kZ_k^\parallel}{Z_k^\parallel} \right)
   T \!\! \sum_{q_4:\,\text{even}}\int_0^\infty \!\!\!\!\! \frac{dw}{
   \Big(w+k^2+q_4^2+\frac{\zetay}{Z_k^\parallel} \Big)^2
   \Big(w+k^2+q_4^2+\frac{\zetax}{Z_k^\parallel} \Big)^2
   } $}
   \notag
   \\
   & \qquad   \qquad 
   - \frac{1}{\pi^2} N_cg^2 k^2\, 
   T\!\!\sum_{q_4:\,\text{odd}}\frac{1}{[q_4^2+E_0(\bar\rho_k)^2]^2}
   \,, 
   \label{eq:fin_dZperp}
   \\
   &{\del_k Z_k^\parallel}   
    =  - \frac{k^2}{\pi^2}
   \frac{\bar\rho_k\big[U''_k(\bar\rho_k)\big]^2}{Z_k^\parallel
   Z_k^\perp}
   \,T\!\! \sum_{q_4:\,\text{even}}\int_0^\infty \!\!\!\!\! \frac{dw}{
   \Big(w+k^2+q_4^2+\frac{\zetay}{Z_k^\parallel} \Big)^2
   \Big(w+k^2+q_4^2+\frac{\zetax}{Z_k^\parallel} \Big)^2
   }
   \notag
   \\
   & \qquad \qquad 
   -\frac{1}{2\pi^2} N_c g^2 |eB|\,T \!\!\sum_{q_4:\,\text{odd}}  
   \sum_{n=0}^{\infty} \frac{\alpha_n}{[q_4^2 + E_n(\bar\rho_k)^2]^2}  
   \,, \label{eq:fin_dZpara}
  \end{align}
  with the definitions 
  \begin{gather}
   U_k' \equiv \del U_k/\del \rho\,, \qquad U_k'' \equiv \del^2 U_k/\del \rho^2\,,
   \\
   \alpha_n \equiv \begin{cases}\ 1~~~(n=0)\\\
		   2~~~(n\geq 1)\end{cases}\!\!\!\!\!,
   \qquad 
   E_n (\rho)\equiv \sqrt{k^2 + 2|eB|n + 2g^2\rho\,}\,, 
   \label{eq:alp-E}
   \\
   E_\pi(\rho)\equiv \sqrt{k^2+\frac{Z_k^\perp p_\perp^2+U_k'(\rho)}{Z_k^\parallel}}\,,
   \quad 
   E_\sigma(\rho)\equiv \sqrt{k^2+\frac{Z_k^\perp p_\perp^2+U_k'(\rho)+
   2\rho U_k''(\rho)}{Z_k^\parallel}}\,, 
   \label{eq:EpiEsig}
   \\
   \bar\rho_k\equiv
   \underset{\rho>0}{\mathrm{argmin}}\big\{U_k(\rho)-h\sqrt{2\rho}\,\big\}\,,
   \\
   \zetay \equiv U'_k(\bar\rho_k)\,, \qquad  
   \zetax \equiv U'_k(\bar\rho_k)+2\bar\rho_k U''_k(\bar\rho_k)\,, 
   \label{eq:bare_masses}
   \\
   \sum_{q_4:\,\text{odd}} \equiv  
   \underset{q_4=(2\ell +1)\pi T}{\sum_{\ell =-\infty}^\infty} \,,\qquad  
   \sum_{q_4:\,\text{even}} \equiv  
   \underset{q_4=2\ell \pi T}{\sum_{\ell=-\infty}^\infty}\,.
  \end{gather}
  The meson masses \eqref{eq:bare_masses} are \emph{bare masses}, which
  should not be confused with the renormalized (physical) masses
  introduced later in section \ref{sc:pob}.  The two primes $(')$ in
  \eqref{eq:fin_dU} imply that the sum and the integral are divergent;
  we will comment more on this below. As one can see from the presence
  of $\del_kZ_k^\parallel$ in the RHS of \eqref{eq:fin_dU} and
  \eqref{eq:fin_dZperp}, the flow of $U_k$ and $Z_k^\perp$ depend on the
  flow of $Z_k^\parallel$, whereas the flow of $Z_k^\perp$ and
  $Z_k^\parallel$ depend on $U_k$ through $\bar\rho_k$. Thus these three
  coupled equations must be solved simultaneously. We note that
  \eqref{eq:fin_dU} does not agree with the flow equations in
  \cite{Skokov:2011ib, Andersen:2012bq,Andersen:2013swa} even for
  $Z_k^\perp=Z_k^\parallel =1$, because the regulator we use is entirely
  different from those in \cite{Skokov:2011ib,Andersen:2012bq,
  Andersen:2013swa}.  The formulas \eqref{eq:fin_dU},
  \eqref{eq:fin_dZperp} and \eqref{eq:fin_dZpara} can be simplified
  analytically so as to facilitate numerical evaluation; see the
  appendices \ref{sc:der_flow_Uk} and \ref{sc:der_flow_Zk} for details.
  
  Even without relying on numerical analysis, one can understand to some
  extent the dynamics of the system through inspection of these flow
  equations.  The second term in $\del_k U_k(\rho)$, \eqref{eq:fin_dU},
  originates from the fermionic contribution to the flow equation
  (cf.~\eqref{eq:FRG-}). The summation over $n$ manifestly embodies the
  Landau level structure of fermion's energy levels, and the lowest
  ($n=0$) Landau level becomes dominant in a strong magnetic field. The
  fact that the prefactor which is normally $k^4$
  \cite{Schaefer:2004en,Skokov:2010wb,Herbst:2010rf} is now replaced by $k^2|eB|$ in
  \eqref{eq:fin_dU} implies that the dynamics of fermions in a strong
  magnetic field is effectively reduced to $(1+1)$-dimensions.  This
  illustrates how the \emph{dimensional reduction}
  \cite{Gusynin:1994xp,Gusynin:1995nb} in the fermionic sector takes
  place.
  
  What is more nontrivial is the dimensional reduction \emph{in the
  bosonic sector} \cite{Fukushima:2012kc}. In our FRG setup, the only
  source of anisotropy of meson dynamics is the asymmetry between
  $\del_k Z_k^\perp$ and $\del_k Z_k^\parallel$. An important difference
  between them is that $\del_k Z_k^\perp$ has no explicit dependence on
  $eB$ in contrast to $\del_k Z_k^\parallel$; one can anticipate that
  this feature will make $Z_k^\perp$ less sensitive to $eB$ than
  $Z_k^\parallel$, which turns out to be true as demonstrated in section
  \ref{sc:num_res}. Another notable difference is that the fermionic
  contribution in \eqref{eq:fin_dZperp} is multiplied by $k^2$ whereas
  that in $\del_k Z_k^\parallel$ is multiplied by $|eB|$.  This means
  that the growth of $Z_k^\parallel$ toward $k=0$ should be enhanced in
  a strong magnetic field, while no such effect is present for
  $Z_k^\perp$. These two characteristics of $\del_k Z_k^\perp$ and
  $\del_k Z_k^\parallel$ provide a rough understanding on how and why
  the magnetic field induces anisotropy in the propagation of neutral
  mesons.
    
    \paragraph{Taylor expansion method}
    In order to make the flow equation numerically more tractable, we
    expand the effective potential as a polynomial around the minimum:
    \begin{align}
     \label{eq:U'_}
     U_k (\rho) & = \sum_{n=0}^{2} a_k^{(n)}\frac{(\rho-\bar\rho_k)^n}{n!}\,,
     \\
     \bar\rho_k & \equiv \underset{\rho}{\rm argmin}\,\big\{ U_k(\rho) - h\sqrt{2\rho} \,\big\}\,.
    \end{align}
    Note that $a_k^{(1)}$ is nonzero since $\bar\rho_k$ is not a minimum
    of $U_k (\rho)$. The expansion up to second order in $\rho$ is
    normally sufficient to describe a second-order phase transition
    \cite{Stokic:2009uv}. Then the flows of $a_k^{(1)}$ and $a_k^{(2)}$
    are easily found as
    \begin{align}
     \del_k a_k^{(1)} = \frac{ \del_k U'_k\Big|_{\bar\rho_k} }
     {\displaystyle 1+\frac{(2\bar\rho_k)^{3/2}}{h}a_k^{(2)} }
     \qquad \text{and} \qquad 
     \del_k a_k^{(2)} =  \del_k U''_k\Big|_{\bar\rho_k} \,,
     \label{eq:taylor_flow}
    \end{align}
    while $\bar\rho_k$ is determined from the relation $\displaystyle
    \bar\rho_k=\frac{h^2}{2\big(a_k^{(1)}\big)^2}$ at each step of the
    flow.  (The flow of $a_k^{(0)}$ is simply ignored as it plays no
    dynamical role.) One can derive $\del_k U'_k$ and $\del_k U''_k$
    from \eqref{eq:fin_dU} by taking derivatives with respect to $\rho$
    (see the appendix \ref{sc:taylorU} for final expressions). The flow
    equations for $Z_k^\perp$ and $Z_k^\parallel$ are readily obtained
    from \eqref{eq:fin_dZperp} and \eqref{eq:fin_dZpara} upon
    substitution of \eqref{eq:U'_}. Now the problem reduces to solving
    coupled ordinary differential equations for five variables:
    $\bar\rho_k$, $a_k^{(1)}$, $a_k^{(2)}$, $Z_k^\perp$ and
    $Z_k^\parallel$.

    \paragraph{Problem of UV renormalization}
    It is intriguing to observe that the UV divergence encountered in
    \eqref{eq:fin_dU} \emph{disappears} once we take the derivative of
    $\del_k U_k$ with $\rho$\,: both the integral and the sum are
    convergent. This means that the UV divergence only appears in the
    constant term of $U_k(\rho)$.  Therefore, within the Taylor
    expansion scheme described above, \emph{no} UV cutoff is necessary
    to make the flows of $a_k^{(1)}$ and $a_k^{(2)}$ finite! The full
    expressions of $\del_k U'_k\big|_{\bar\rho_k}$ and $\del_k
    U''_k\big|_{\bar\rho_k}$ obtained without UV cutoff are lengthy and
    are presented in the appendix \ref{sc:taylorU}.

    In principle one could also argue that an explicit UV cutoff has to
    be applied because the quark-meson model is after all a
    \emph{low-energy} effective model of QCD. To assess the sensitivity
    of infrared observables to the UV regularization scheme, we have
    also solved the flow equations \emph{with} an explicit UV cutoff
    $\sim$ 1 GeV and compared the obtained results with those from the
    cutoff-free scheme.  We found that while quantitative differences
    are present, the global tendencies of results from both schemes are
    the same, including the monotonic increase of $T_{\rm pc}$ as a
    function of $eB$. Therefore we will only present the numerical
    results obtained within the \emph{cutoff-free} scheme in the next
    section.
    
    \paragraph{LPA and mean-field approximation}
    Finally, let us comment on other related schemes. In LPA we ignore
    nontrivial scale dependence of the propagators, which amounts to
    setting $Z_k^\perp=Z_k^\parallel\equiv1$ in \eqref{eq:fin_dU}.  This
    approximation has been employed to study chiral models in a magnetic
    field \cite{Skokov:2011ib,Andersen:2012bq,Andersen:2013swa}.
    
    The conventional mean-field approximation is attained from our flow
    equation by setting bosonic fields to their expectation values and
    removing the bosonic loop contribution in \eqref{eq:fin_dU}
    altogether.  The resulting flow equation now reads
    \begin{align} 
     \del_k U_k(\rho) 
     & =  - \frac{1}{2\pi^2}N_ck^2  |eB|\ {\sum_{n=0}^{\infty}}'
     \frac{\alpha_n}{E_n(\rho)}\tanh \frac{E_n(\rho)}{2T} \,.
    \end{align}
    It is instructive to integrate both sides over $k$ explicitly:
    \begin{align}
     U_{k=0}(\rho)  & = U_{k=\Lambda}(\rho) - \int_{0}^{\Lambda}dk \left[
     - \frac{1}{2\pi^2}N_ck^2  |eB|\ {\sum_{n=0}^{\infty}}'
     \frac{\alpha_n}{E_n(\rho)}\tanh \frac{E_n(\rho)}{2T}\right]
     \\
     & = U_{k=\Lambda}(\rho) + 4 N_c T \frac{|eB|}{2\pi}\ {\sum_{n=0}^{\infty}}' \alpha_n 
     \int_{0}^{\Lambda}\frac{dk}{2\pi}~k \frac{\del}{\del k}\left(\log
     \cosh \frac{E_n(\rho)}{2T}\right)
     \\
     & = U_{k=\Lambda}(\rho) - N_c \frac{|eB|}{2\pi}\ {\sum_{n=0}^{\infty}}' \alpha_n 
     \int_{-\Lambda}^{\Lambda}\frac{dp_3}{2\pi}\big[ 
     E_n(\rho) + 2T \log(1+\ee^{-E_n(\rho)/T})  
     \big]\,, \label{eq:U_MF}
    \end{align}
    where in the last step we have discarded an irrelevant constant and
    a surface term resulting from partial integration, and relabelled
    $k$ as $p_3$ so that $E_n(\rho)=\sqrt{p_3^2+2|eB|n+2g^2\rho\,}$ can
    be interpreted as the energy of a quark in the $n$-th Landau
    level. As claimed above, \eqref{eq:U_MF} reproduces the
    thermodynamic potential in the mean-field approximation
    \cite{Mizher:2010zb,Gatto:2010pt}.  The expectation value of $\rho$
    should be determined from the minimization of
    $U_{k=0}(\rho)-h\sqrt{2\rho}$\,.

  \subsection{Physical quantities}
  \label{sc:pob} Let us define physical quantities attained in the $k\to
  0$ limit of the flow equation.  The essence is that the minimum of the
  effective potential gives the condensate $\langle \sigma \rangle$
  while the curvature around the minimum gives the meson masses. In the
  presence of the field renormalization, however, these quantities are
  nontrivially renormalized and care must be taken in comparing results
  from FRG with those from other methods, such as lattice simulations.
  In this subsection we wish to spell out the notations and definitions
  of all observables we consider, as a preparation for section
  \ref{sc:num_res} where they are evaluated by numerically solving the
  flow equation.
  
  Firstly, the dynamical quark mass is given by
  \begin{align}
    M_q \equiv gf_\pi^{\rm bare} = g\sqrt{2\bar\rho_{k=0}}\,,
   \label{eq:def_of_M_q}
  \end{align}
  where $f_\pi^{\rm bare}=\langle\sigma\rangle$ is the \emph{bare} pion
  decay constant. ($\langle\sigma\rangle>0$ for $h>0$.)
  
  Next, we note that the dispersion of the mesons follows from
  \eqref{eq:TB} via analytic continuation as
  \begin{align}
    Z_k^{\parallel}p_0^2 - Z_k^\perp (p_1^2+p_2^2) 
    - Z_k^{\parallel}p_3^2 - \zetasp^2 = 0\,, 
    \label{eq:disp}
  \end{align}
  with the \emph{bare} masses $\zetasp$ defined in
  \eqref{eq:bare_masses}.  Thus the screening mass in the directions
  orthogonal to the magnetic field (i.e., the \emph{transverse}
  screening mass), $m_{\sigma,\pi}^{\perp}$\,, and the screening mass
  along the direction of the magnetic field (i.e., the
  \emph{longitudinal} screening mass), $m_{\sigma,\pi}^{\parallel}$\,,
  are given by
  \begin{gather}
   m_{\sigma,\pi}^{\perp} \equiv \frac{\zetasp}{\sqrt{Z^\perp_{k=0}}} \qquad 
    \text{and} \qquad 
    m_{\sigma,\pi}^{\parallel} \equiv
   \frac{\zetasp}{\sqrt{Z^\parallel_{k=0}}} \,, 
\label{eq:def_of_screening_masses}
  \end{gather}
  respectively. The \emph{pole mass} is equal to
  $m_{\sigma,\pi}^{\parallel}$ within our effective action. It is also
  evident from \eqref{eq:disp} that the transverse velocity $v_\perp$ of
  mesons (i.e., the velocity of mesons in the directions perpendicular
  to the magnetic field) is given by \footnote{Strictly speaking,
  $v_\perp$ in \eqref{eq:v_def} is equal to the transverse velocity of
  mesons only when $m_{\sigma,\pi}=0$. However we stick to calling this
  quantity the velocity for brevity.}
  \begin{align}
    \label{eq:v_def}
    v_\perp \equiv \sqrt{\frac{Z_{k=0}^\perp}{Z_{k=0}^\parallel}}  \,.
  \end{align}
  It has been suggested in model calculations that $v_\perp^2\ll 1$ in a
  strong magnetic field \cite{Gusynin:1994xp,Gusynin:1995nb,
  Fayazbakhsh:2012vr,Fukushima:2012kc,Fayazbakhsh:2013cha} and it is one
  of our aims to check this at finite temperature in the framework of
  FRG, incorporating the effect of fluctuations of interacting mesons.
  
  Interestingly, in the presence of a magnetic field the decay constant
  of the neutral pion also exhibits anisotropy
  \cite{Fayazbakhsh:2013cha}.  This is due to the fact that the coupling
  of pions to the axial vector current is direction-dependent in a
  magnetic field.  Although the definition of a `decay constant' in a
  thermal media is nontrivial (see e.g.,
  \cite{Pisarski:1996mt,Son:2001ff,Son:2002ci}), following
  \cite{Jungnickel:1995fp,Berges:1997eu} we shall define the
  \emph{transverse} and \emph{longitudinal} pion decay constants at
  finite temperature by
  \begin{align}
    f_\pi^\perp \equiv \sqrt{Z_{k=0}^\perp\,}\,f_{\pi}^{\rm bare}
    & = \sqrt{2 Z_{k=0}^\perp \bar\rho_{k=0}}  \quad \text{and}
    \label{eq:def_of_pion_decay_constants_1}
    \\
    f_\pi^\parallel \equiv \sqrt{Z_{k=0}^\parallel\,}\,f_{\pi}^{\rm bare}
    & = \sqrt{2 Z_{k=0}^\parallel \bar\rho_{k=0}} \,,
   \label{eq:def_of_pion_decay_constants_2}
  \end{align}
  respectively. This convention is motivated by the fact that the chiral
  effective Lagrangian of the neutral pion to lowest order assumes a
  particularly simple form
  \begin{align}
    \mathcal{L}_{\rm eff} = \frac{f_\pi^{\perp 2}}{4}(\del_\perp U)^2 
    + \frac{f_\pi^{\parallel 2}}{4}(\del_\parallel U)^2 +\dots\,,
  \end{align}
  where $U(x)$ is a $\U(1)$ field whose phase describes the pion,
  $\del_\perp=(\del_1, \del_2)$ and $\del_\parallel=(\del_3, \del_4)$.
  In the limit of a weak magnetic field, $f_\pi^\perp/f_\pi^\parallel
  \to 1$ and $\mathcal{L}_{\rm eff}$ reduces to the familiar form.
  
  This completes the formulation of FRG for the quark-meson model.

 \section{Numerical results}
 \label{sc:num_res} 
 
 In this section we will show results of integrating the flow equations
 numerically. In order to estimate the impact of mesonic fluctuations,
 we will contrast results from three approximations: LPA plus
 scale-dependent wave function renormalizations (which we term ``full
 FRG''), LPA, and the mean-field approximation.
 
 One of our purposes is to understand the phase structure from the
 viewpoint of chiral symmetry. After describing the initial conditions
 of the flow in section \ref{sc:parafix}, we will present results for
 the constituent quark mass ($M_q$) at finite temperature and magnetic
 field in section \ref{sc:tpc}.  From the temperature dependence of
 $M_q$ the pseudo-critical temperature of the chiral phase transition is
 estimated and its dependence on the magnetic field is examined.

 The neutral meson dynamics acquires anisotropy in an external magnetic
 field through the quark loop contributions.  The second purpose of our
 FRG analysis is to see the anisotropy of neutral meson modes.  In
 section \ref{sec:meson-modes-under}, we calculate some observables such
 as meson screening masses, and examine their directional dependence at
 finite temperature and external magnetic field.
  
  \subsection{Parameter fixing}
  \label{sc:parafix} 
  
  We numerically solved the Taylor-expanded flow \eqref{eq:taylor_flow}
  with the second-order Runge-Kutta method (RK2) for full FRG, LPA, and
  the mean-field approximation, respectively. The initial scale of the
  RG flow is fixed at 600 MeV. In LPA and the mean-field approximation,
  we have four initial parameters: $a_{k= \Lambda}^{(1)}$,
  $a_{k=\Lambda}^{(2)}$, $h$ and $g$. In the full FRG calculation, in
  addition, we need to specify initial values for the wave function
  renormalizations, $Z^{\perp}$ and $Z^{\parallel}$. All those initial
  conditions are gathered in Table~\ref{tb:inicond}.
   \begin{table}[tb]
   \centering
    \begin{tabular}{|c||c|c|c|c|c|c|c|c|}
     \hline
     &  $N_c$ & $g$ & $\Lambda$ & $h/\Lambda^3$  
     & $a_{k=\Lambda}^{(1)^{\mathstrut}}/\Lambda^2$ 
     & $a_{k=\Lambda}^{(2)}$ 
     & $Z_{k=\Lambda}^\parallel$ & $Z_{k=\Lambda}^\perp$ 
     \\\hline\hline 
     Full FRG & 3 & 2.0 & 600 & 0.00596 & 0.489 & 1.0 
     & 0.002 & 0.236 \\
     \hline
     LPA & 3 & 2.76 & 600 & 0.00835 & 0.732 & 5.0 
     & --- & --- \\\hline
     Mean field & 3 & 2.76 & 600 & 0.00820 & 0.947 & 0.25 
     & --- & --- 
     \\\hline
    \end{tabular}
    \caption{Initial conditions for the flow equation at $k=\Lambda$. The
    column for $\Lambda$ is given in MeV.}  
    \label{tb:inicond}
    \vspace{\baselineskip}
    \centering
    \begin{tabular}{|c||c|c|c|c|c|c|}
     \hline
     &\multicolumn{5}{|c|}{Observables at $k=0$}& \multirow{2}{*}{$T_{\rm pc} $}\\\cline{2-6}
     & $f_\pi $ & $m_\pi$ & $m_{\sigma}$  & $M_{q}$ &
     $Z_{k=0}^{\parallel,\perp}$ & \\  \hline\hline
     Full FRG & $93.4^*$ & $138^*$ & $411^*$ & 257 & 0.529 &178
     \\\hline
     LPA  & 94.2 & 138 & 407 & 260 & --- &194
     \\\hline
     Mean field  & 92.5 & 138 & 417 & 261 & --- &174
     \\\hline
    \end{tabular}
    \caption{Resulting physical values at $T=3$ MeV and
    $eB=0.5m_{\pi}^2$ and the pseudo-critical temperature $T_{\rm pc}$
    at $eB=0.5m_{\pi}^2$. The columns for $f_\pi$, $m_\pi$, $m_\sigma$,
    $M_q$ and $T_{\rm pc}$ are given in MeV.  The values with \mbox{star
    $^*$} ($f_\pi$, $m_\pi$ and $m_\sigma$ for Full FRG) are obtained
    after the wave function renormalization (see section \ref{sc:pob}).}
    \label{tb:physical values}
  \end{table}
  
  In Table~\ref{tb:physical values}, resulting physical values at $k=0$
  are shown for each approximation at \mbox{$T=3$ MeV} and
  $eB=0.5m_{\pi}^2$.  (We checked that observables hardly vary for $0
  \lesssim eB\lesssim 0.5m_{\pi}^2$\,, so $eB=0.5m_{\pi}^2$ is small
  enough to be considered as the limit of vanishing magnetic field.)
  The initial flow parameters were tuned in each approximation so as to
  reproduce physical values for $M_q, m_{\pi}, m_{\sigma}$ and
  $f_{\pi}$.  This makes our model a good laboratory for QCD in the real
  world.  As explained in section~\ref{sc:pob}, the values of physical
  observables in the full FRG calculation ($m_{\pi}$, $m_{\sigma}$ and
  $f_{\pi}$) are subject to the wave function renormalization.
  
  In vacuum ($T=eB=0$), the Euclidean $\SO(4)$ symmetry is
  intact. However this is not automatically realized in our setup due to
  the fact that the regulators used here (\eqref{eq:RB_3d} and
  \eqref{eq:RF_3}) break the $\SO(4)$ symmetry \emph{explicitly},
  regardless of the magnetic field strength and temperature.  Indeed
  $\del_k Z_k^\perp$ in \eqref{eq:fin_dZperp} does not agree with
  $\del_k Z_k^\parallel$ in \eqref{eq:fin_dZpara} even in the vacuum
  limit ($T,\,eB\to 0$). We cure this problem by fine-tuning the initial
  conditions $Z^{\perp,\parallel}_{k=\Lambda}$ so that
  $Z^{\perp}_{k=0}=Z^{\parallel}_{k=0}$ holds at \mbox{$T=3$ MeV} and
  $eB=0.5m_{\pi}^2$.  This is how $Z_{k=\Lambda}^{\perp,\parallel}$ in
  Table \ref{tb:inicond} are fixed. We have used the same set of initial
  values at all temperatures.\footnote{The deviation of
  $Z_{k=0}^\perp/Z_{k=0}^\parallel$ from $1$ turns out to be at most
  $10\%$ over the range \mbox{$0<T<360$ MeV} at $eB=0.5m_{\pi}^2$.}
  
  In Table~\ref{tb:physical values}, we also summarize the
  pseudo-critical temperature ($T_{\rm pc}$) in each approximation
  scheme at \mbox{$eB=0.5m_{\pi}^2$}.  Here $T_{\rm pc}$ is determined
  from the peak of the temperature derivative of the constituent quark
  mass. In the following subsections, we shall normalize the temperature
  axis of every plot by $T_{\rm pc}$ at $eB=0.5m_{\pi}^2$ to facilitate
  comparison of the three approximations.
    
  \subsection{Pseudo-critical temperature}
  \label{sc:tpc}
  
  \begin{figure}[t!]
   \begin{center}
    \includegraphics[width=0.5\columnwidth]{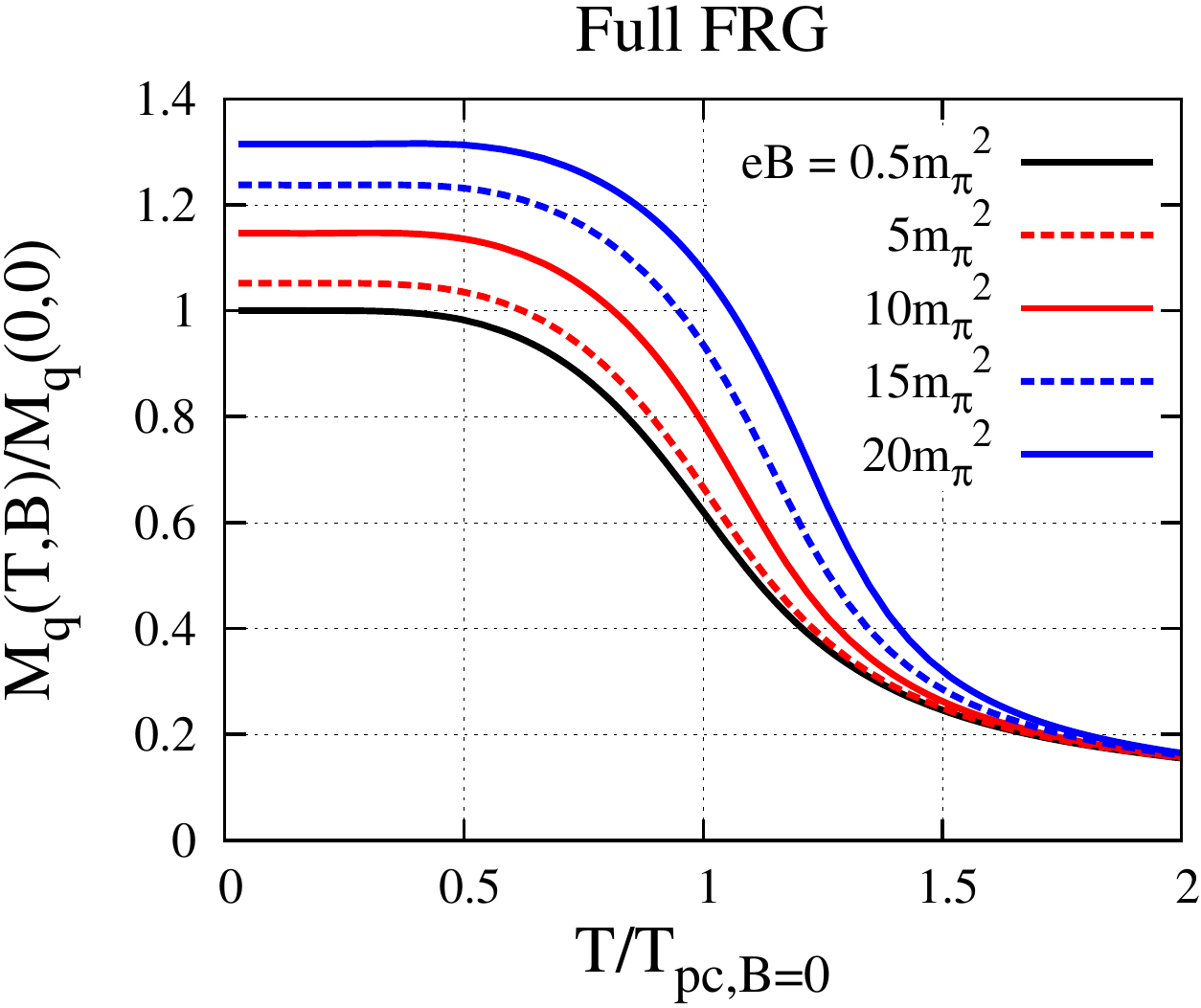}
   \end{center}
   \vspace{-.5\baselineskip}
   \begin{minipage}{0.5\hsize}
    \begin{center}
     \includegraphics[width=1.0\columnwidth]{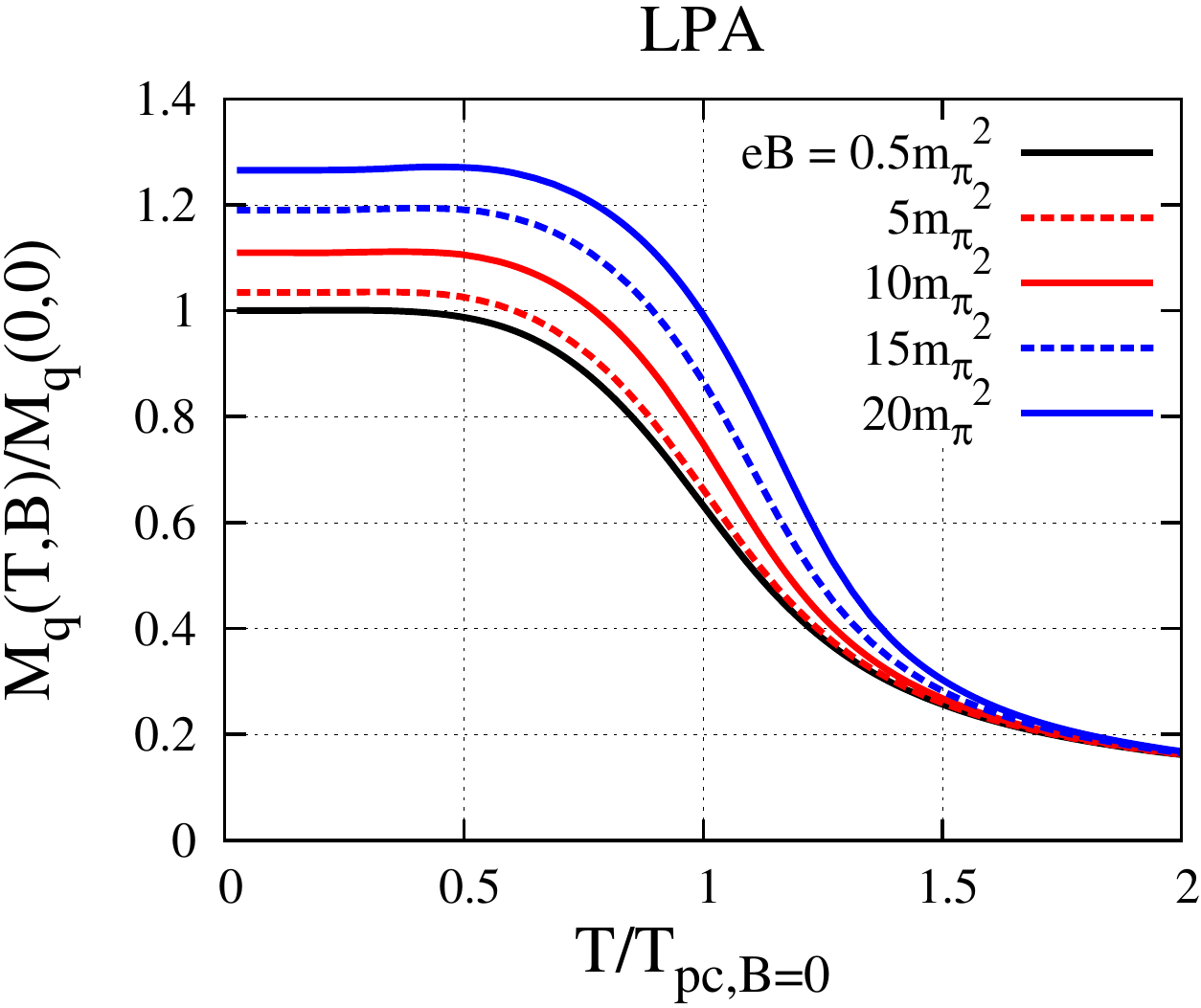}
    \end{center}
   \end{minipage}
   \begin{minipage}{0.5\hsize}
    \begin{center}
     \includegraphics[width=1.0\columnwidth]{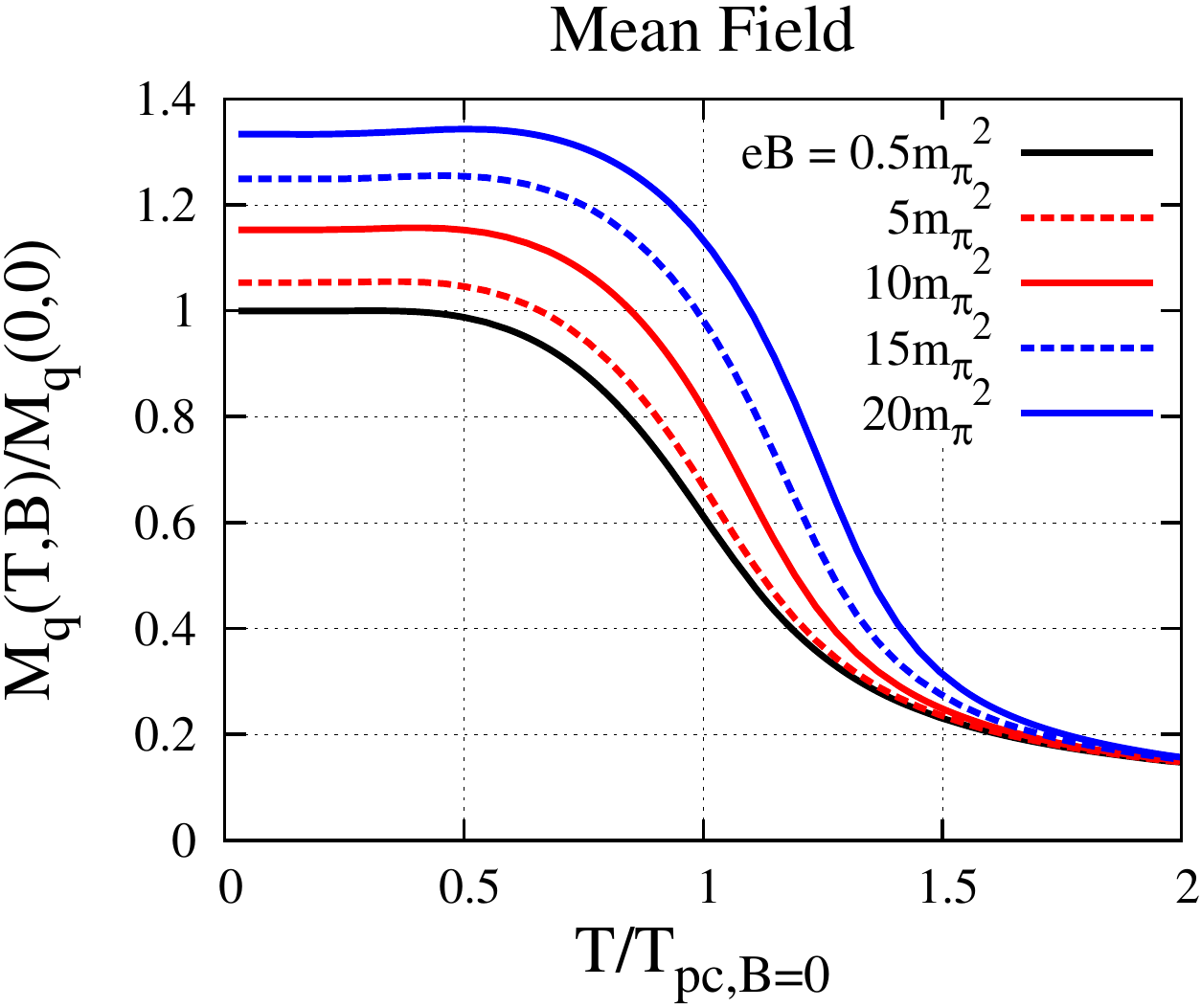}
    \end{center}
   \end{minipage}
   \caption{The constituent quark mass at finite temperature and
   magnetic field from full FRG (top), LPA (bottom, left) and the
   mean-field approximation (bottom, right). The vertical axis is
   normalized to $1$ at $T=eB=0$.
   }  
   \label{fig:1_Mq_vs_T_full}  
  \end{figure}

  The constituent quark mass $M_q$ is proportional to the bare pion
  decay constant~(cf.~(\ref{eq:def_of_M_q})) and serves as an order
  parameter for the chiral symmetry breaking. In
  Fig.~\ref{fig:1_Mq_vs_T_full}, we show the temperature dependence of
  $M_q$ in full FRG, LPA, and the mean-field approximation, with varying
  external magnetic field.  The three plots share the same qualitative
  features. At low temperature, chiral symmetry is spontaneously broken
  and quarks acquire a mass of order $300$ MeV. At high temperature,
  chiral symmetry is effectively restored: the dynamical mass drops to
  around 15\% of the vacuum value at $T=2T_{\rm pc}$.  Since quarks have
  the current mass, $M_q$ never reaches zero even above $T_{\rm pc}$.

  From Fig.~\ref{fig:1_Mq_vs_T_full} one can read off the external
  magnetic field dependence of the constituent quark mass. In all the
  three approximations, $M_q$ increases monotonically with $|eB|$ at all
  temperatures below $2T_{\rm pc}$.  This behavior, called magnetic
  catalysis, has been observed in lattice simulations
  \cite{D'Elia:2012tr} as well as in various chiral effective models
  \cite{Shovkovy:2012zn}. The increase of $M_q$ with $|eB|$ is slower in LPA 
  than in the mean-field approximation, which is attributable to the
  meson-loop contribution to the flow of $U_k$ that counteracts the
  symmetry breaking effect of fermions.  On the other hand, our new
  result from full FRG, which also includes effects of the wave function
  renormalization, turns out to be closer to the mean-field
  approximation than LPA.

  \begin{figure}[t!]
   \begin{center}
    \includegraphics[width=0.5\columnwidth]{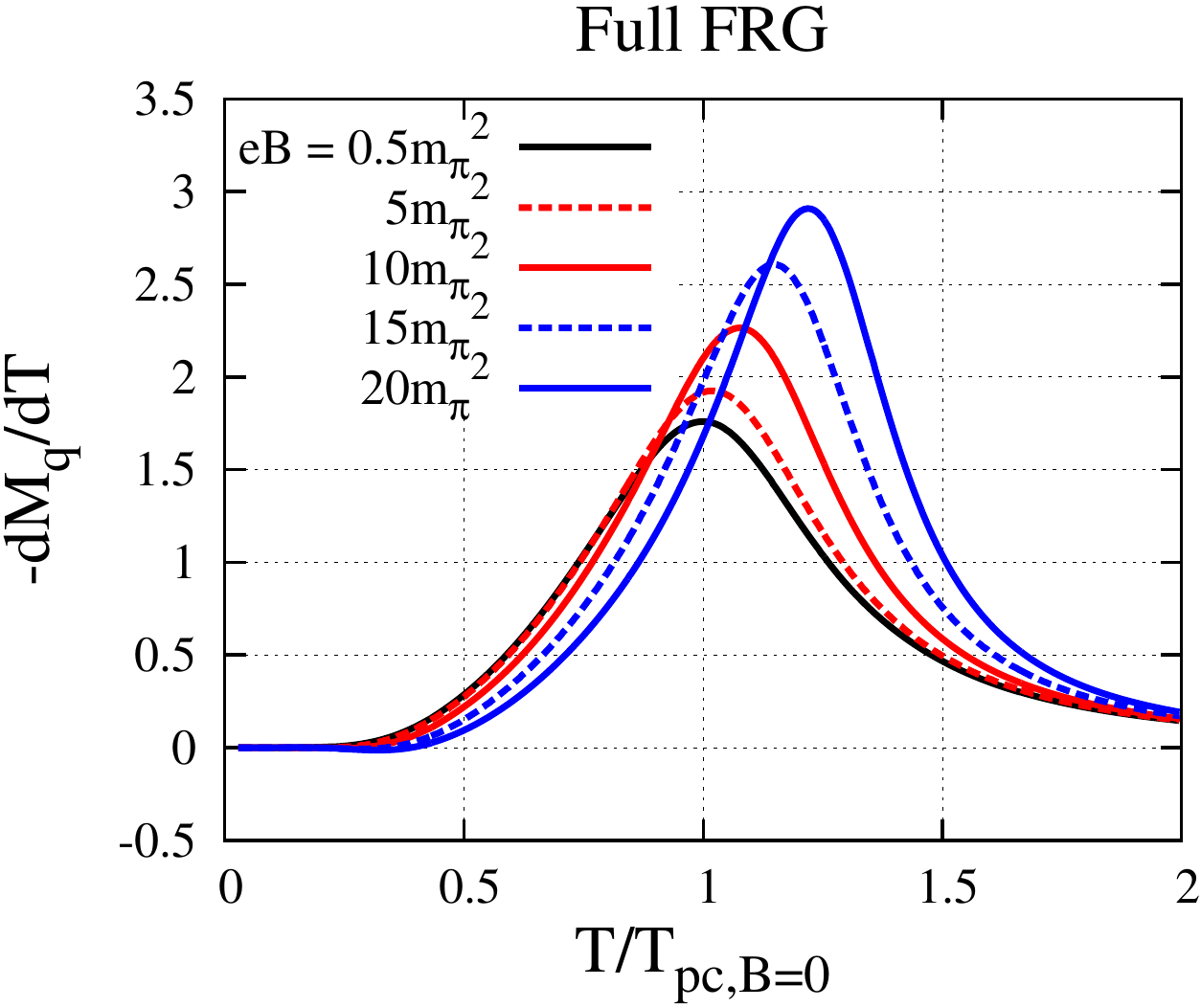}
   \end{center}
   \vspace{-.5\baselineskip}
   \begin{minipage}{0.5\hsize}
    \begin{center}
     \includegraphics[width=1.0\columnwidth]{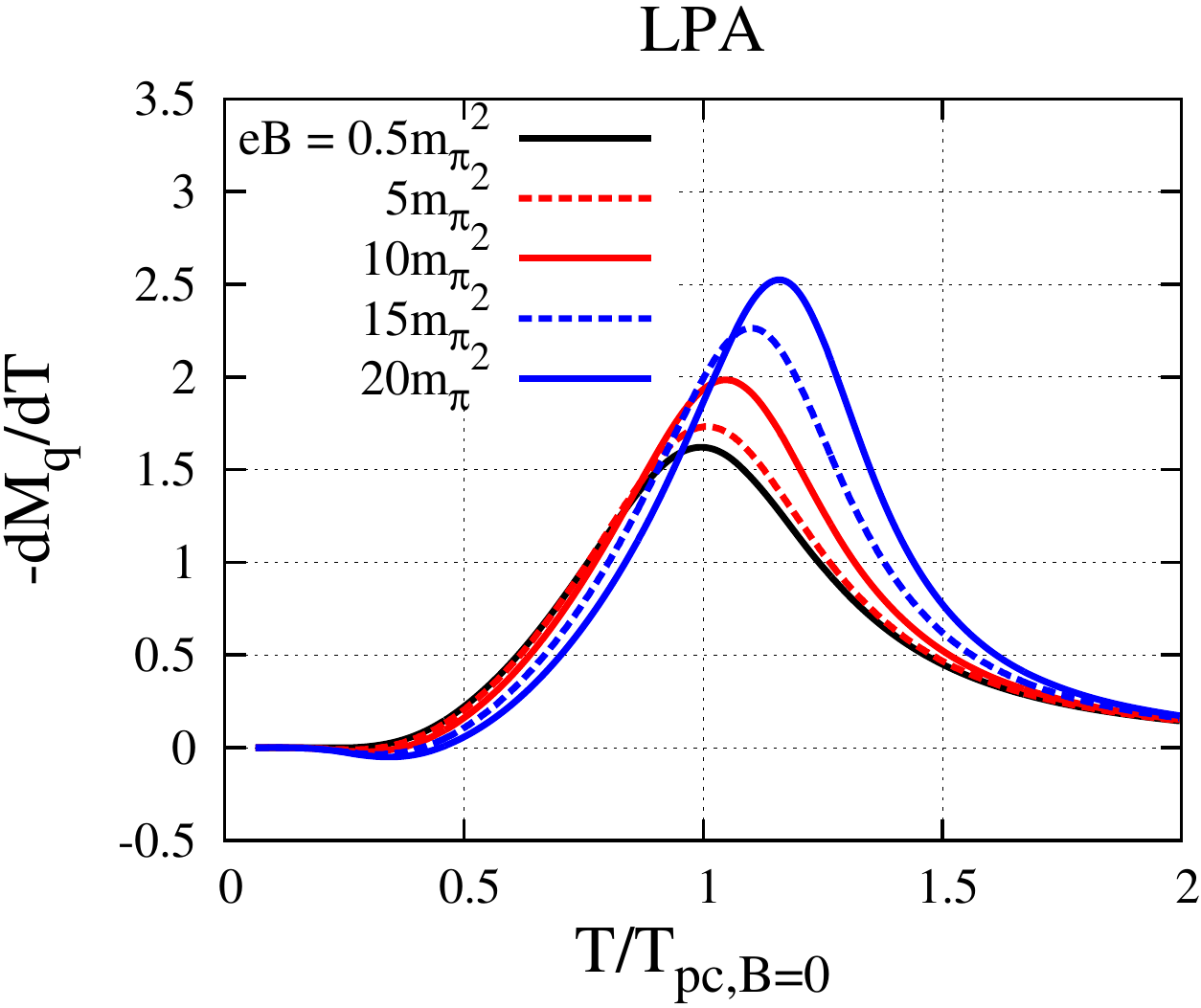}
    \end{center}
   \end{minipage}
   \begin{minipage}{0.5\hsize}
    \begin{center}
     \includegraphics[width=1.0\columnwidth]{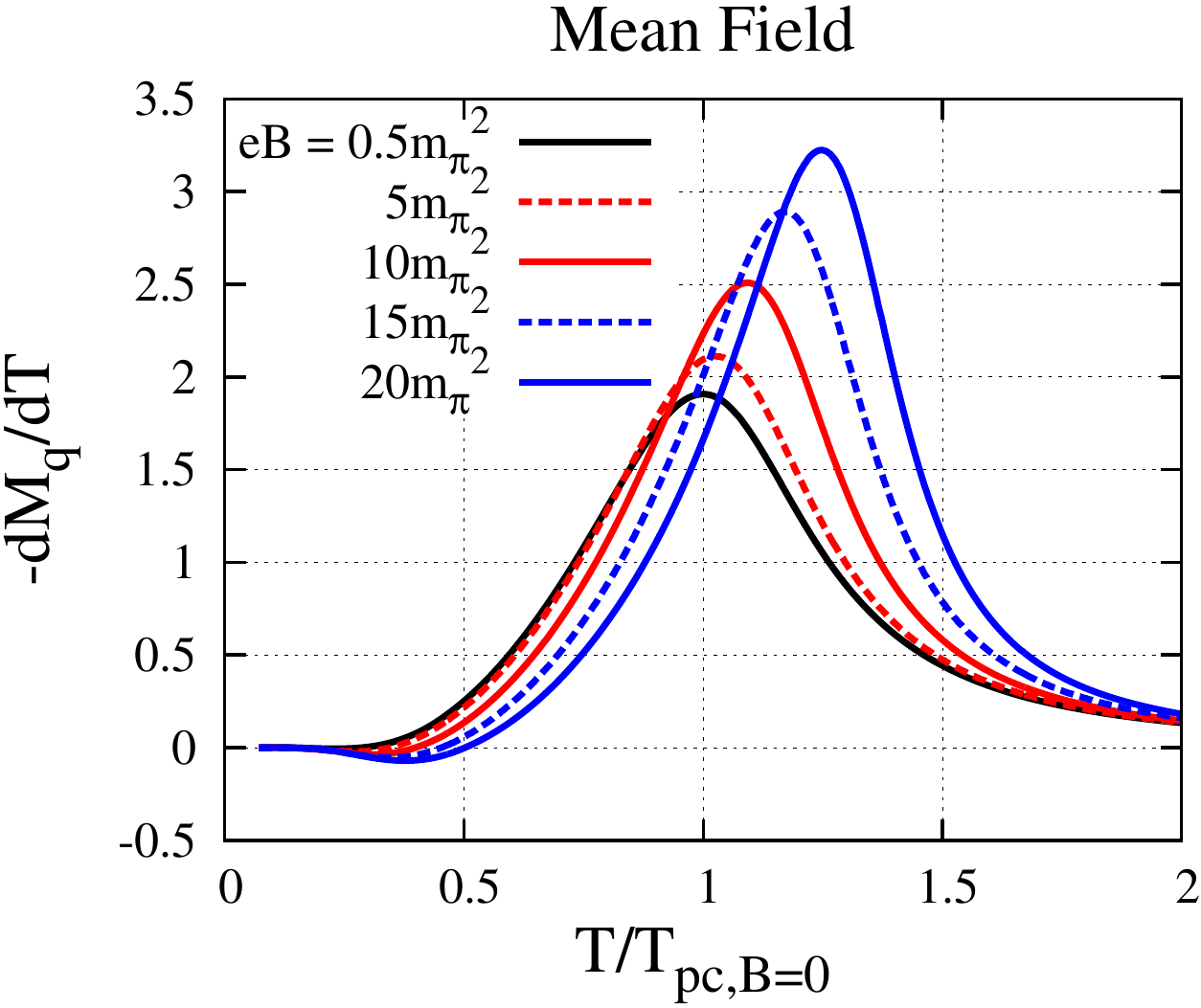}
    \end{center}
   \end{minipage}
   \caption{The slope of the constituent quark mass at finite
   temperature and magnetic field from full FRG (top), LPA (bottom,
   left) and the mean-field approximation (bottom, right).}
   \label{fig:2_Mq_slope_vs_T}
  \end{figure}
  
  In Fig.~\ref{fig:2_Mq_slope_vs_T} we show the temperature derivative
  of $M_q$ for various values of the external magnetic field.  The peaks
  of these curves define the pseudo-critical temperature, $T_{\rm pc}$.
  Clearly, in all approximations, the peak temperature moves to a higher
  value for a stronger magnetic field. This tendency is consistent with
  many other works based on chiral effective models. However this is at
  odds with the recent lattice QCD calculation with light quarks
  \cite{Bali:2011qj,Bali:2012zg}.  The plots in
  Fig.~\ref{fig:2_Mq_slope_vs_T} suggest that the inclusion of the wave
  function renormalization alone does not resolve the discrepancy
  between the lattice QCD and chiral effective models.
  
  \begin{figure}[t!]
   \begin{center}
    \includegraphics[width=0.5\columnwidth]{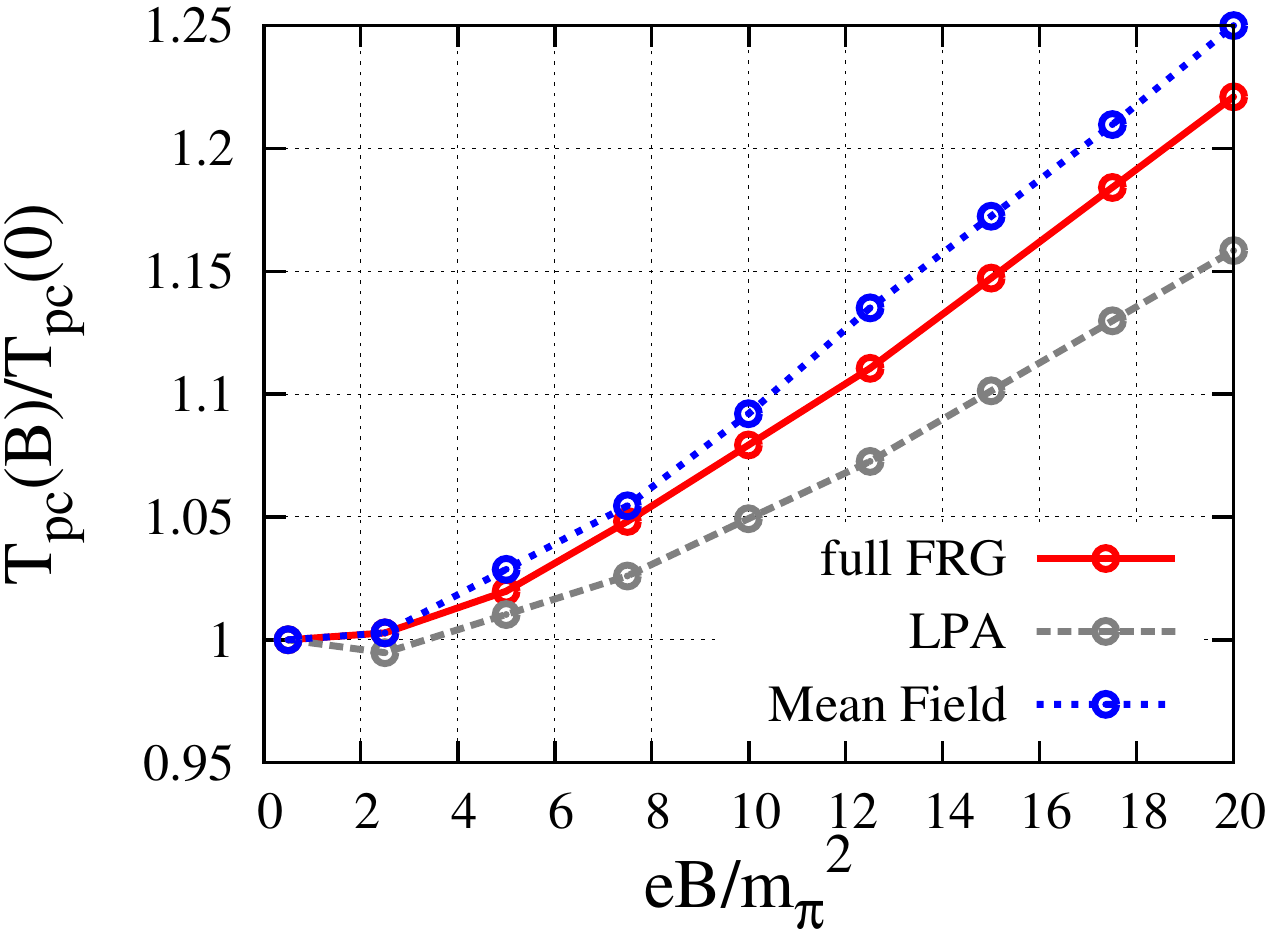}
   \end{center}
   \vspace{-1.5\baselineskip}
   \caption{Magnetic field dependence of the pseudo-critical
   temperatures from three approximations.}
   \label{fig:Tc_vs_B}
  \end{figure}
  In Fig.~\ref{fig:Tc_vs_B}, we plot the pseudo-critical temperature
  versus magnetic field (in units of $m_\pi^2$) for each
  approximation. In all the three cases $T_{\rm pc}$ rises monotonically
  with $|eB|$, and $T_{\rm pc}$ in LPA and full FRG shows a milder
  increase than $T_{\rm pc}$ in the mean-field approximation, owing to
  the effect of mesonic fluctuations. 
  This tendency is in discord with the previous work
  with two light flavors \cite{Skokov:2011ib}, where $T_{\rm pc}$ of LPA 
  showed a stronger increase than that of the mean field. 
  We speculate that the difference comes from the absence of the charged pions 
  in our work. 
  
  Figure \ref{fig:Tc_vs_B}, somewhat unexpectedly, also shows that
  $T_{\rm pc}$ from full FRG rises \emph{more steeply than} $T_{\rm pc}$
  of LPA and behaves like that of the mean-field approximation.  In the
  next subsection we will try to give a possible explanation to this
  trend based on the pion pole mass behavior at finite temperature.

  \subsection{Meson modes under magnetic field}
  \label{sec:meson-modes-under} 
  
  In the last subsection we discussed the dynamical quark mass and the
  chiral restoration temperature. In what follows, we will present and
  discuss results related to the meson properties. The neutral mesons
  change their nature under strong external magnetic field because they
  are made of charged quarks. The most prominent feature is an
  anisotropy of the neutral meson modes. To investigate this issue in a
  quantitative manner we have calculated various observables related to
  the anisotropy of the neutral meson modes.
  
  \begin{figure}[t!]
   \begin{minipage}{0.5\hsize}
    \begin{center}
     \includegraphics[width=1.00\columnwidth]{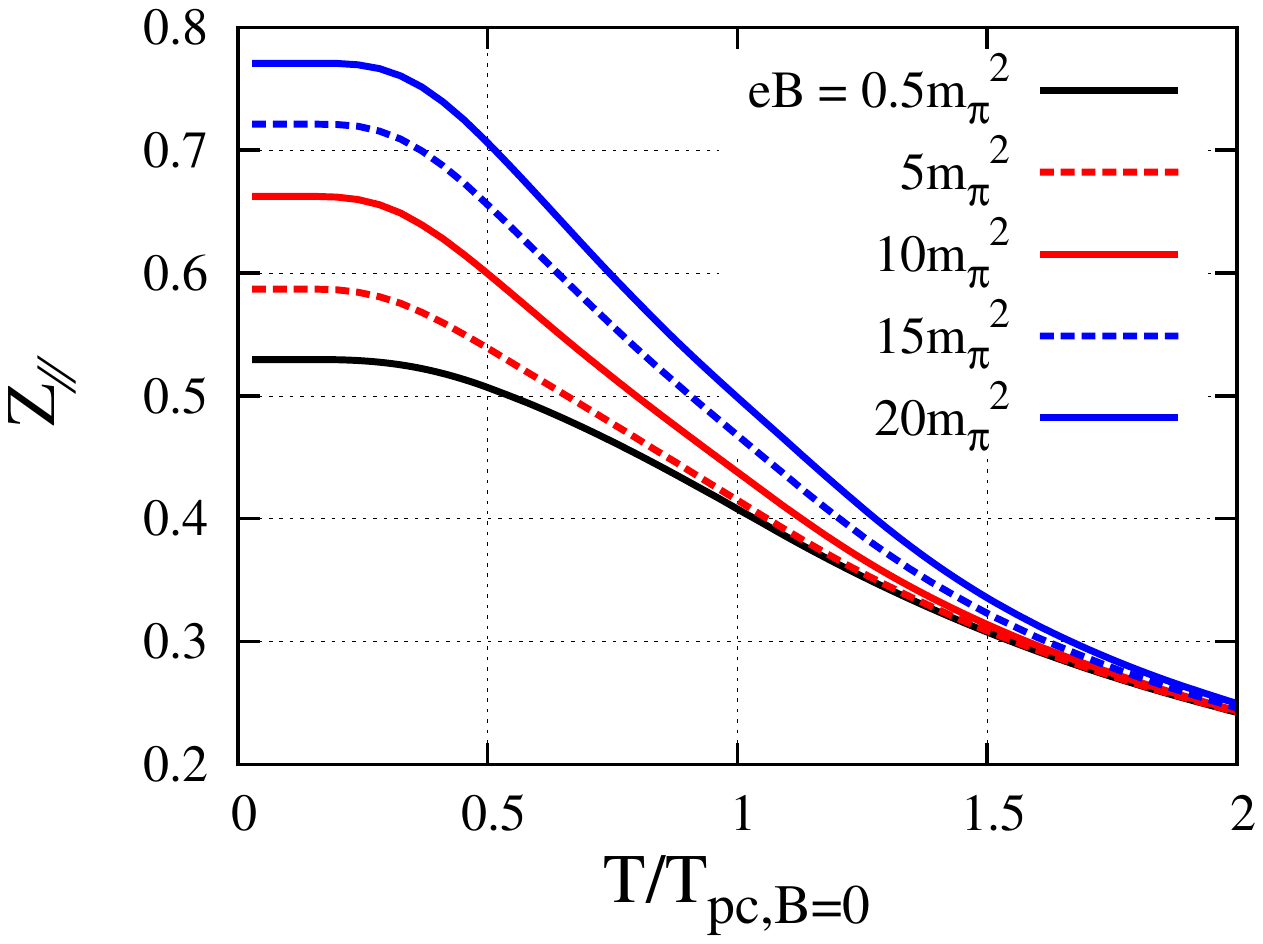}
    \end{center}
   \end{minipage}
   \begin{minipage}{0.5\hsize}
    \begin{center}
     \includegraphics[width=1.00\columnwidth]{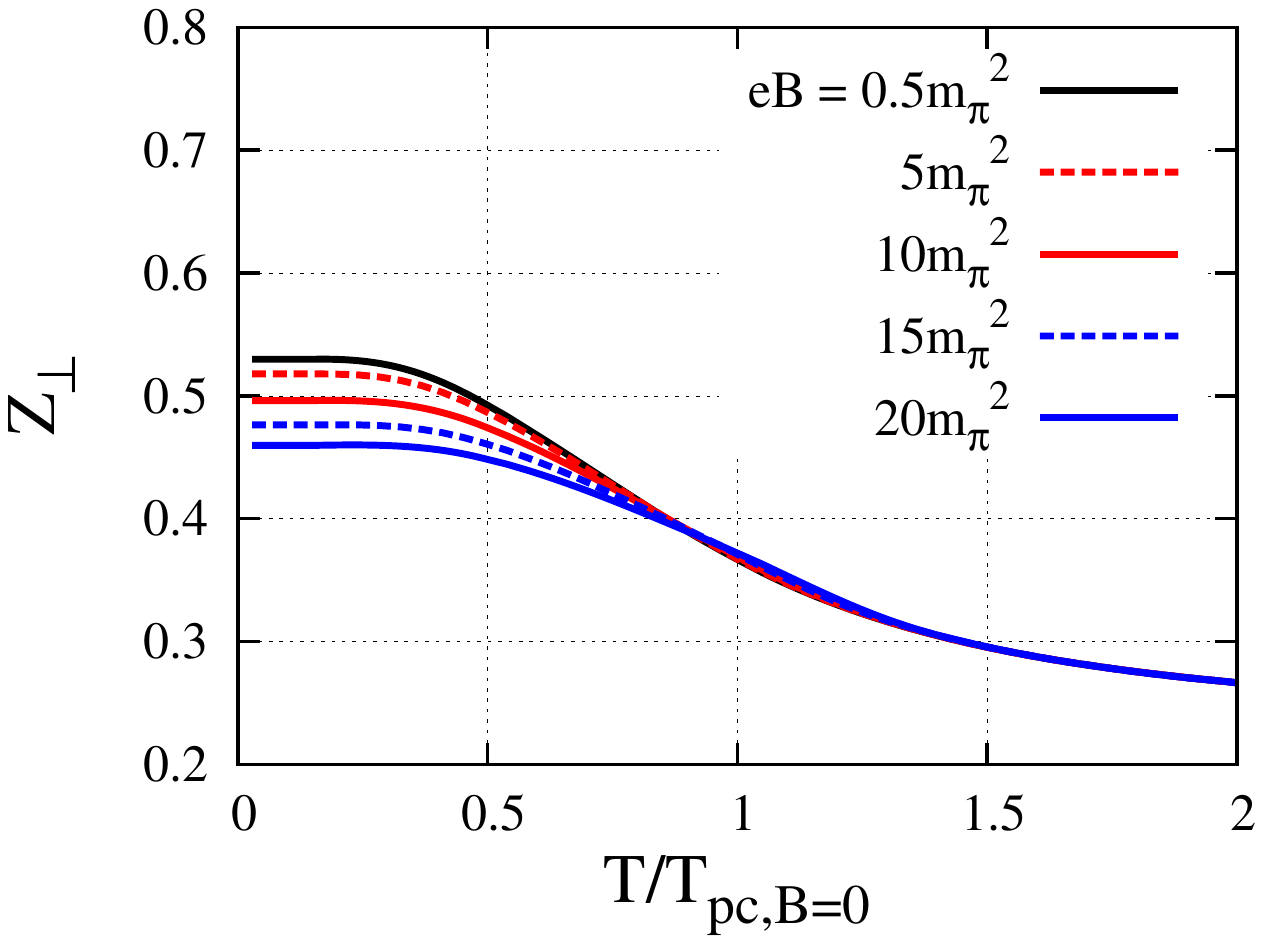}
    \end{center}
   \end{minipage}
   \vspace{-0.5\baselineskip}
   \caption{Parallel (left) and perpendicular (right) wave function
   renormalization factors of mesons.}  \label{fig:10_Z_vs_T}
  \end{figure}

  Let us begin with the wave function renormalization factors, which are
  the most central objects in our beyond-LPA analysis.  In
  Fig.~\ref{fig:10_Z_vs_T} we show $Z^{\parallel}$ and $Z^{\perp}$ at
  finite temperature and external magnetic field.  There one can observe
  several marked features:
  \begin{description}
    \setlength{\itemsep}{-1.5mm}
    \item[~~(a)] At high temperature, both $Z^{\parallel}$ and $Z^{\perp}$ diminish substantially and 
    become insensitive to the magnetic field. 
    \item[~~(b)] $Z^{\parallel}$ \emph{increases} sharply with $|eB|$. 
    \item[~~(c)] By contrast, $Z^{\perp}$ \emph{decreases} with $|eB|$. However 
    $Z^{\perp}$ shows only weak dependence on $|eB|$ at all temperatures.  
  \end{description}
  These features can be understood, at least qualitatively, from the
  flow equations in \eqref{eq:fin_dZperp} and
  \eqref{eq:fin_dZpara}. First of all, we remark that the meson
  contributions to $\del_k Z_k^{\parallel}$ and $\del_k Z_k^{\perp}$ are
  suppressed at all temperatures, except for the vicinity of $T_{\rm
  pc}$. (We have checked this explicitly by numerically integrating the
  flow equation.) The reason is as follows.  In the meson loop diagram
  (cf.~Fig.~\ref{fig:dg_frg_2pf}), both $\sigma$ and $\pi$ are
  circulating around the loop. Since $\sigma$ is always heavy (except
  near $T_{\rm pc}$) and $\pi$ also gets heavy at high temperature, the
  meson loop contribution turns out to be always suppressed as compared
  to the fermion loop contribution. Therefore the flows of
  $Z^{\parallel}$ and $Z^{\perp}$ are mostly dominated by the fermionic
  contributions in \eqref{eq:fin_dZperp} and \eqref{eq:fin_dZpara}. Now
  we are ready to interpret {\bf (a)}--{\bf (c)} above.
  
  At high temperature, fermions acquire a large screening mass $q_4 \sim
  \pi T$ due to the antiperiodic boundary condition along the $x^4$
  direction. Then the fermionic contribution to \eqref{eq:fin_dZperp}
  and \eqref{eq:fin_dZpara} is strongly suppressed and consequently
  $Z^{\parallel}_k$ and $Z^{\perp}_k$ almost cease to flow. Indeed,
  $Z^\perp_{k=0}\simeq 0.265$ at $T/T_{\rm pc}=2$, which is close to the
  initial value, $Z^\perp_{k=\Lambda}=0.236$. Thus we expect that both
  $Z^{\parallel}$ and $Z^{\perp}$ tend to their initial values at
  sufficiently high temperature. This should be true in a magnetic
  field, too, as long as $\sqrt{eB}$ does not exceed the screening scale
  $\sim \pi T$. This is an intuitive explanation to {\bf (a)}.
  
  As for {\bf (b)}, the increase of $Z^\parallel$ is most likely
  attributable to the enhancement of the lowest Landau level $(n=0)$
  contribution in \eqref{eq:fin_dZpara}.  The contribution from the
  higher Landau levels is clearly suppressed for large $|eB|$ and they
  decouple from the flow of $Z_k^\parallel$.
  
  Let us finally turn to {\bf (c)}. The weak dependence of $Z^{\perp}$
  on the magnetic field, in stark contrast to $Z^\parallel$, is quite
  natural in view of the fact that the flow of $Z^{\perp}$,
  \eqref{eq:fin_dZperp}, has no explicit dependence on $|eB|$.  (This
  fact itself is a result of complicated nontrivial cancellations of
  $|eB|$-dependence among infinite series, as demonstrated in the
  appendix \ref{sc:z_perp_f}.) The slight decrease of $Z^{\perp}$ as a
  function of $|eB|$ is more subtle; we speculate that this tendency
  originates from the enhancement of the constituent quark mass in a
  magnetic field (cf.~Fig.~\ref{fig:1_Mq_vs_T_full}).  Because
  $\bar\rho_k$ grows with $|eB|$ owing to the magnetic catalysis, the
  fermionic contribution in \eqref{eq:fin_dZperp} is suppressed, and the
  growth of $Z_k^{\perp}$ toward $k=0$ is slowed down.  Thus the
  decrease of $Z^\perp_{k=0}$ seems to be a natural consequence of large
  $|eB|$.
  
\begin{figure}[t!]
   \begin{center}
    \includegraphics[width=0.5\columnwidth]{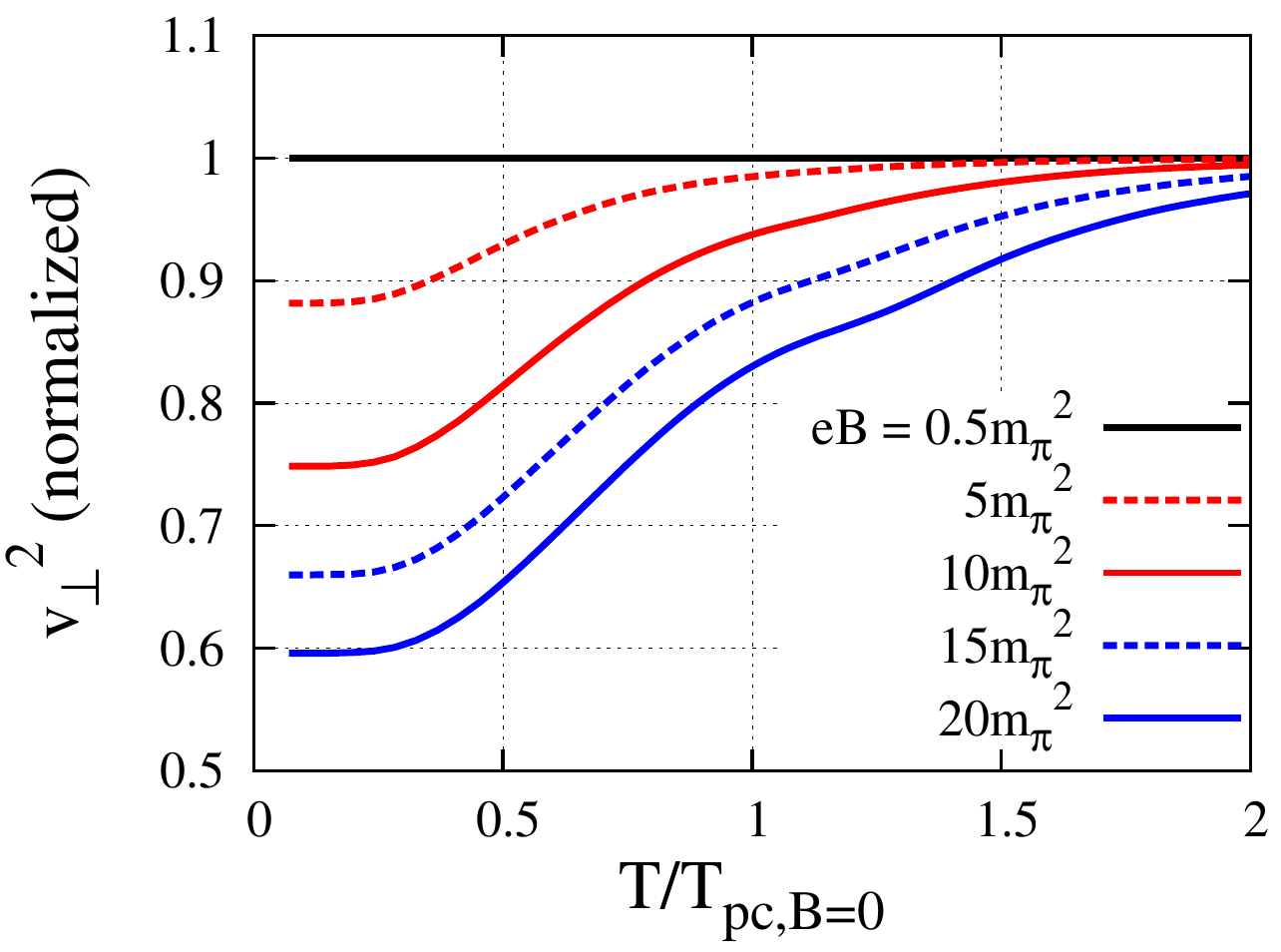}
   \end{center}
   \vspace{-1.5\baselineskip}
   \caption{Squared transverse velocity
   $v_\perp^2=Z^{\perp}/Z^{\parallel}$ with varying external magnetic
   field. The velocity is normalized by that at $eB=0.5m_{\pi}^2$.}
   \label{fig:11_Zs_ratio_vs_T}
   \end{figure}

  The ratio of $Z^{\perp}$ to $Z^{\parallel}$ gives the squared
  transverse velocity, $v_\perp^2$. Even at $eB=0$, $v_\perp^2$ deviates
  from 1 owing to the finite temperature effect.  To see the effect of
  the external magnetic field, it is convenient to normalize
  $v_{\perp}^2$ by that at $eB=0.5m_{\pi}^2$. In
  Fig.~\ref{fig:11_Zs_ratio_vs_T} we show the temperature dependence of
  $v_\perp^2$ thus normalized for varying external magnetic field.  For
  all temperatures, the velocity decreases with $eB$. This behavior is
  consistent with previous works that studied neutral mesons at $T=0$
  \cite{Gusynin:1994xp,Gusynin:1995nb,Fukushima:2012kc}.  Our new
  finding here is that $v_\perp$ has a strong temperature dependence: at
  high temperature $(\gtrsim T_{\rm pc})$ even the magnetic field as
  strong as $20 m_\pi^2$ does not modify $v_\perp^2$ significantly. This
  tendency can naturally be understood by recalling the temperature
  dependence of $Z^\parallel$ and $Z^\perp$ (cf.~{\bf (a)}). Therefore
  the ``dimensional reduction'' of neutral mesons is unlikely to modify
  the nature of the chiral crossover in a qualitative way.

    \begin{figure}[t!]
    \begin{minipage}{0.5\hsize}
    \begin{center}
     \includegraphics[width=1.00\columnwidth]{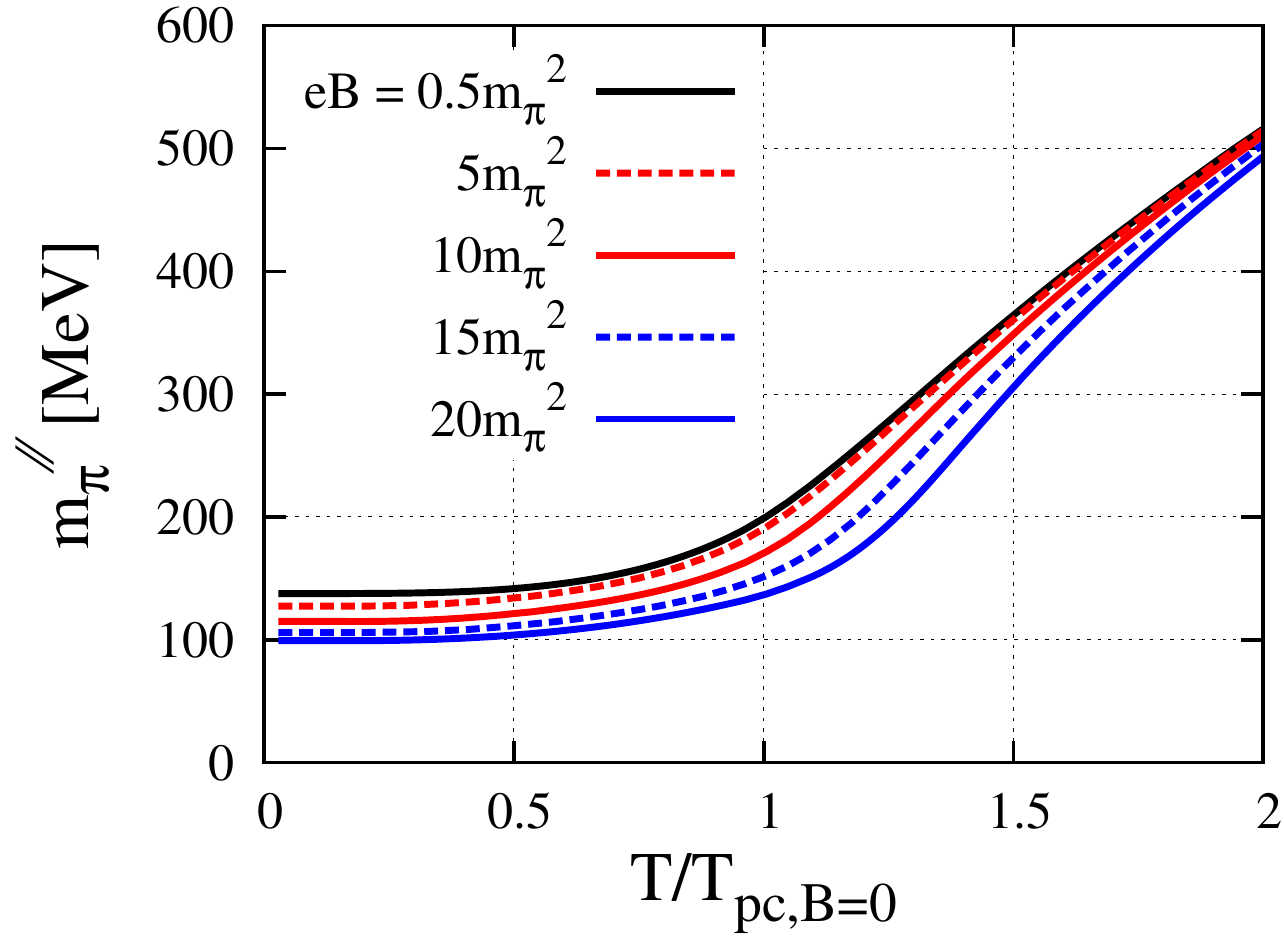}
    \end{center}
   \end{minipage}
   \begin{minipage}{0.5\hsize}
    \begin{center}
     \includegraphics[width=1.00\columnwidth]{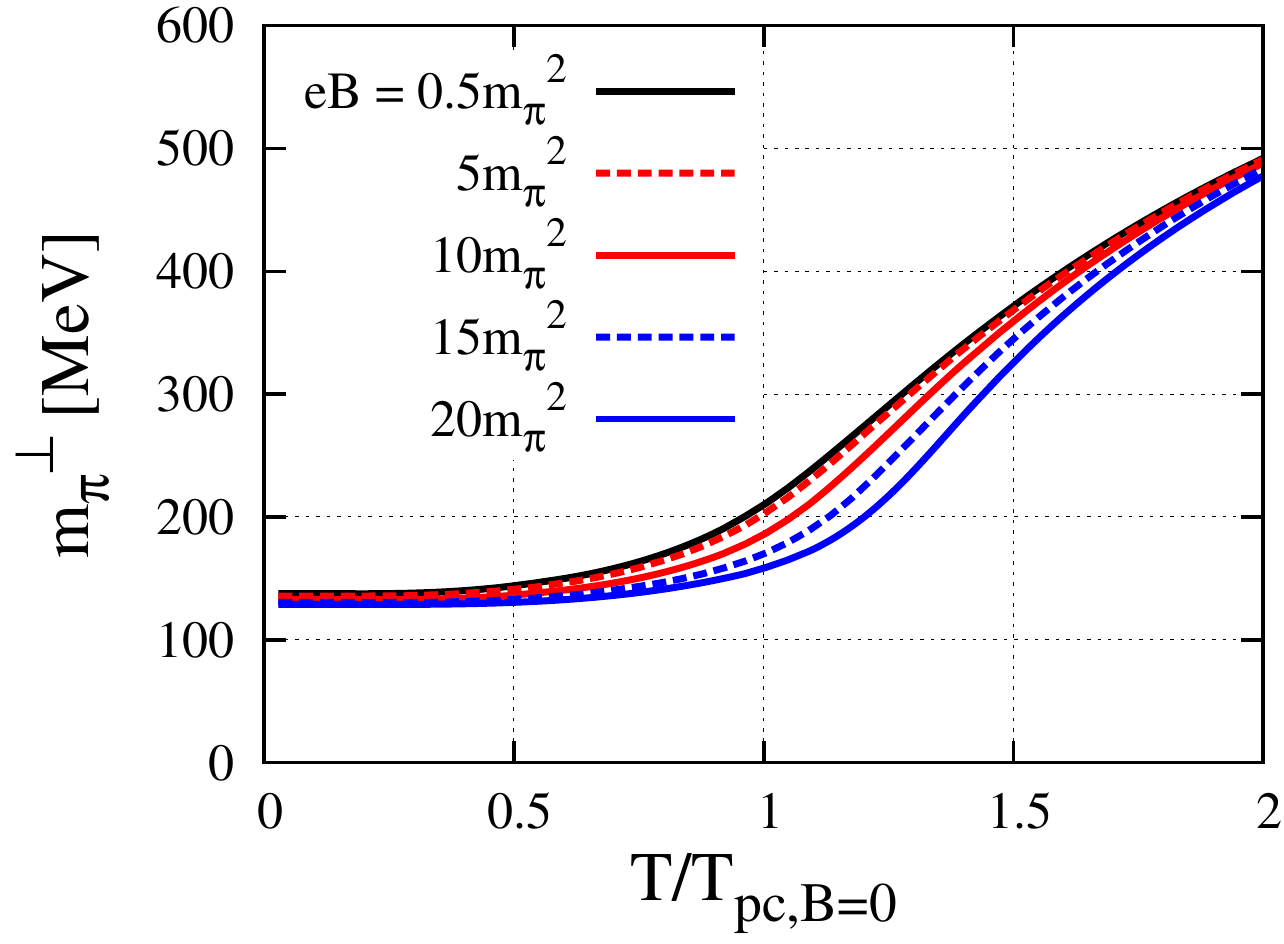}
    \end{center}
    \end{minipage}
    \vspace{-.5\baselineskip}
    \begin{center}
     \includegraphics[width=0.50\columnwidth]{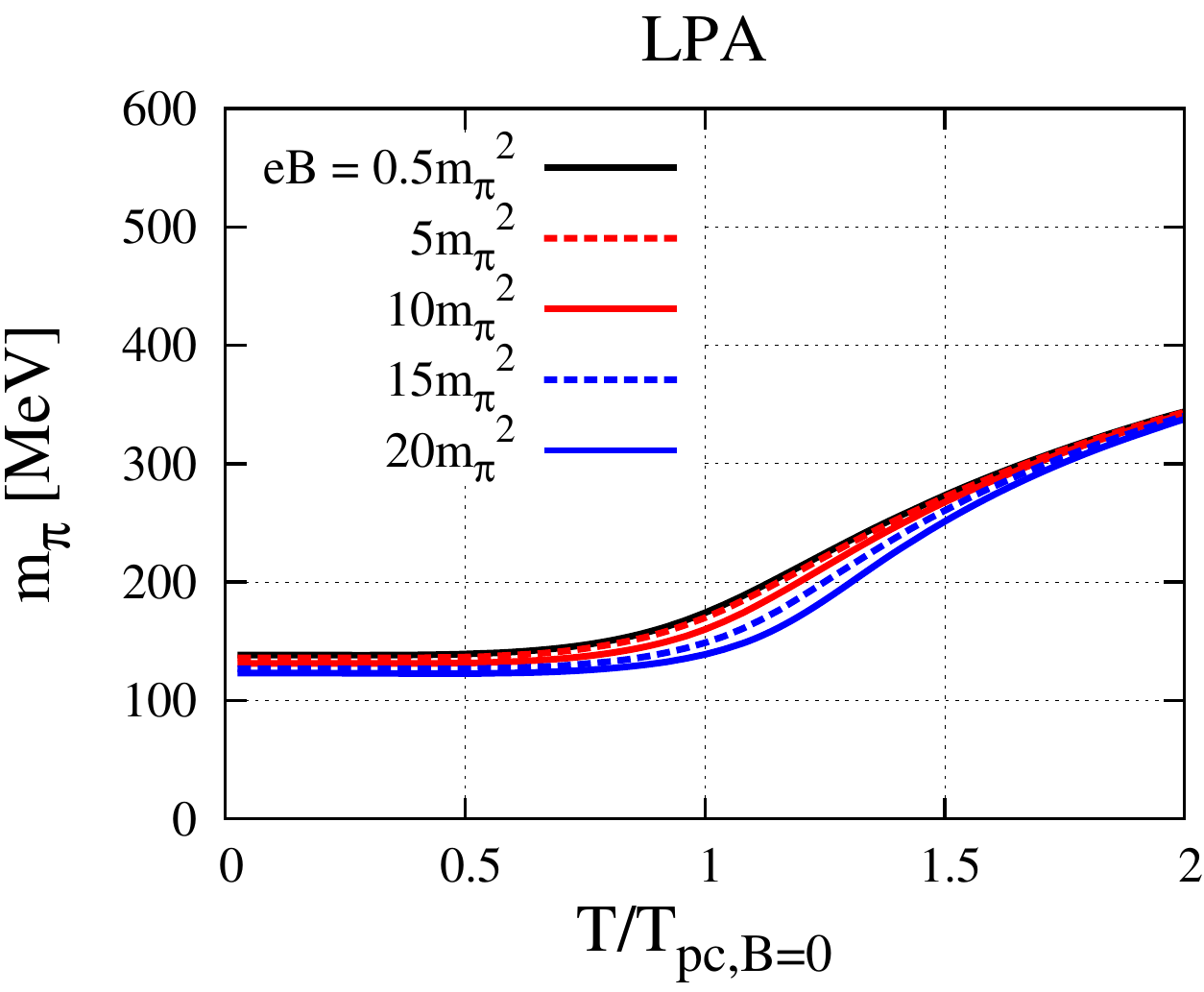}
    \end{center}
    \vspace{-\baselineskip}
   \caption{The longitudinal (top, left) and transverse (top, right)
   pion screening masses from full FRG, and the pion mass from LPA
   (bottom), with varying external magnetic field.}
   \label{fig:pi_mass_full_vs_LPA}
  \end{figure}

  \begin{figure}[t!]
   \begin{minipage}{0.5\hsize}
    \begin{center}
     \includegraphics[width=1.00\columnwidth]{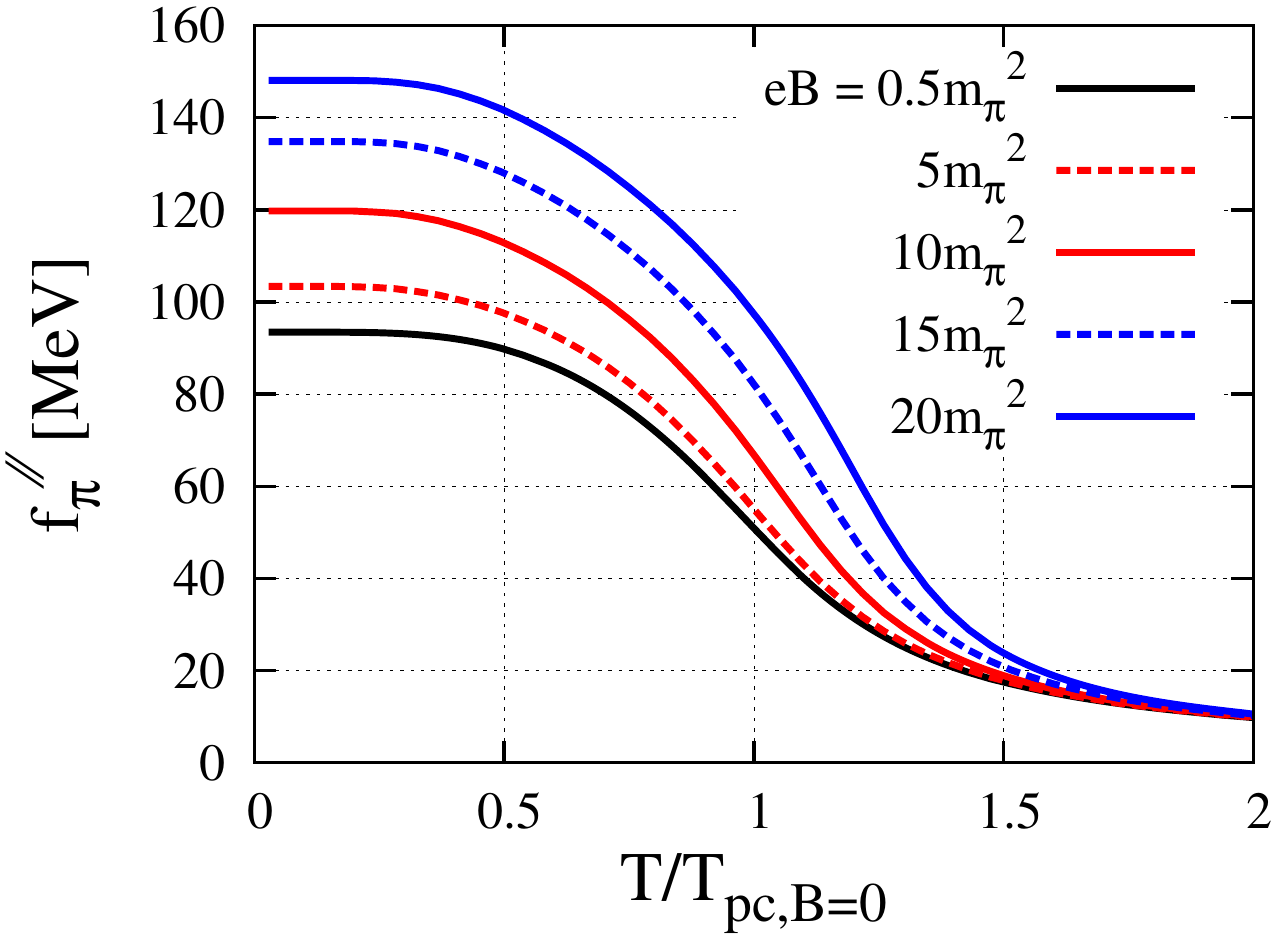}
    \end{center}
   \end{minipage}
   \begin{minipage}{0.5\hsize}
    \begin{center}
     \includegraphics[width=1.00\columnwidth]{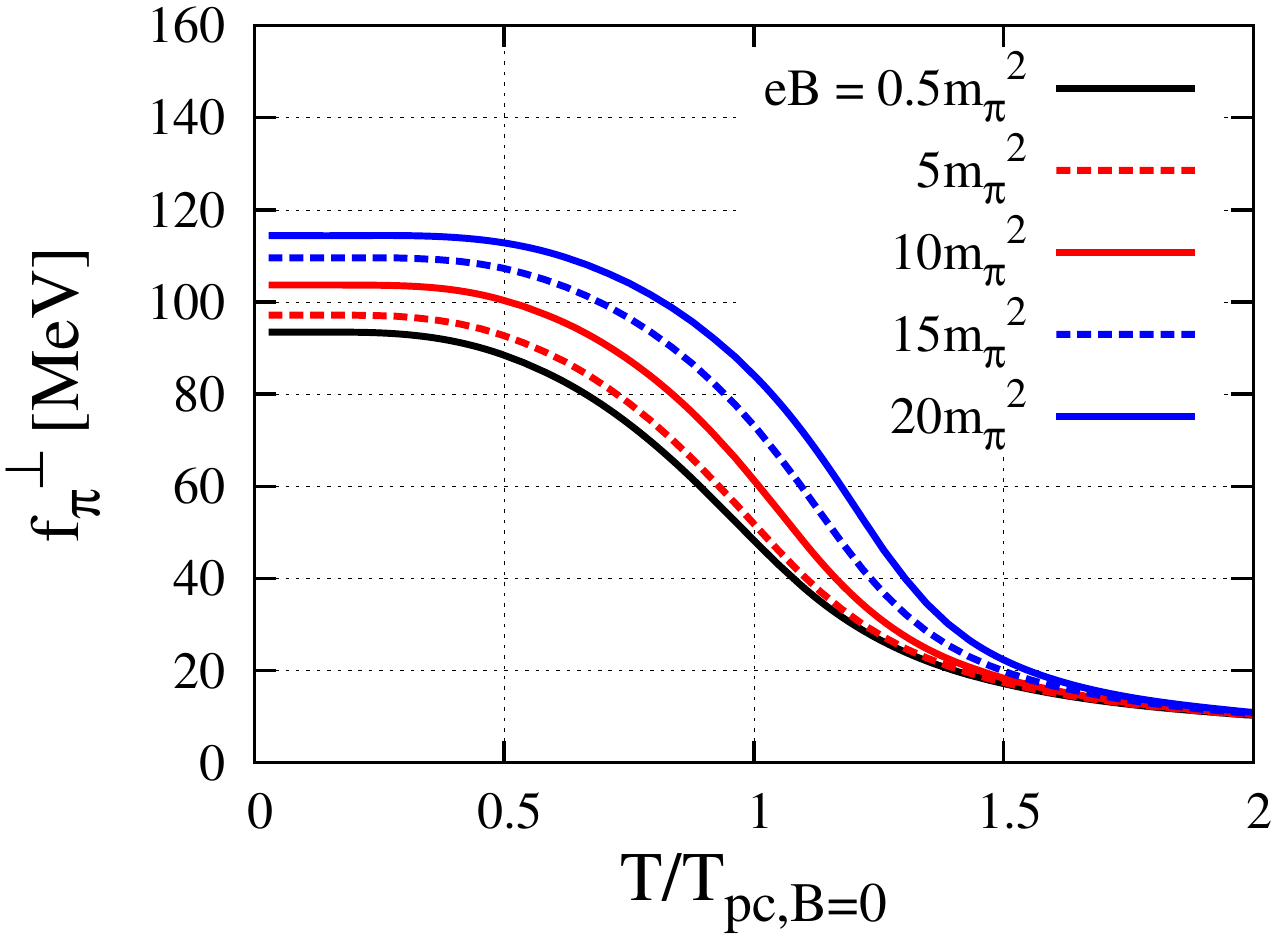}
    \end{center}
   \end{minipage}
   \caption{The longitudinal (left) and transverse (right) pion decay constants for
   varying external magnetic field.}  
   \label{fig:4_f_vs_T}
  \end{figure}
  
  In Fig.~\ref{fig:pi_mass_full_vs_LPA} (top), we show the renormalized
  pion masses obtained in full FRG.  As remarked in section
  \ref{sc:pob}, the screening masses acquire a directional dependence in
  a strong magnetic field.\footnote{Within our truncation the pole
  mass and the longitudinal screening mass are identical, although they
  can be different in QCD at finite temperature.}  For comparison, in
  Fig.~\ref{fig:pi_mass_full_vs_LPA} (bottom) we also present the pion
  mass from LPA. In all three cases, we observe that the neutral pion
  mass \emph{decreases} in a magnetic field. This trend is consistent
  with lattice simulations \cite{Luschevskaya:2012xd,Hidaka:2012mz},
  chiral perturbation theory
  \cite{Shushpanov:1997sf,Agasian:2001ym,Andersen:2012dz,Andersen:2012zc},
  and an analytical study \cite{Orlovsky:2013gha}.
  
  Furthermore, by comparing full FRG with LPA we find that
  $m_\pi^\parallel$ and $m_\pi^\perp$ grow more steeply with $T$ than
  $m_\pi$ in LPA for $T \gtrsim T_{\rm pc}$.  This difference originates
  from the fact that $Z^\parallel$ and $Z^\perp$ decrease rapidly with
  $T$ (cf.~Fig.~\ref{fig:10_Z_vs_T}).  Because of this rapid growth of
  the pion pole mass in full FRG at high $T$, the mesonic contributions
  to the flow are suppressed as compared to LPA.  Therefore it is
  natural that in Fig.~\ref{fig:Tc_vs_B} the pseudo-critical temperature
  of full FRG shows the same trend with the mean-field approximation
  rather than LPA.
  
  In Fig.~\ref{fig:4_f_vs_T}, we present temperature dependence of the
  renormalized longitudinal and transverse pion decay constants (see
  \eqref{eq:def_of_pion_decay_constants_1} and
  \eqref{eq:def_of_pion_decay_constants_2} for their definitions).  At
  each temperature, both pion decay constants increase with $eB$, but
  with different rates.  Because $Z^{\parallel}$ increases with the
  external magnetic field, it enhances the increase of $f_{\pi}^{\rm
  bare}$. On the other hand, $Z^{\perp}$ decreases with the external
  field. Then the increase of $f_{\pi}^{\rm bare}$ is partially canceled
  by $Z_{\perp}$. However the decrease of $Z_{\perp}$ is not rapid
  enough to decrease $f^{\perp}_{\pi}$ with the external magnetic field.

  \begin{figure}[t!]
   \begin{minipage}{0.5\hsize}
    \begin{center}
     \includegraphics[width=1.00\columnwidth]{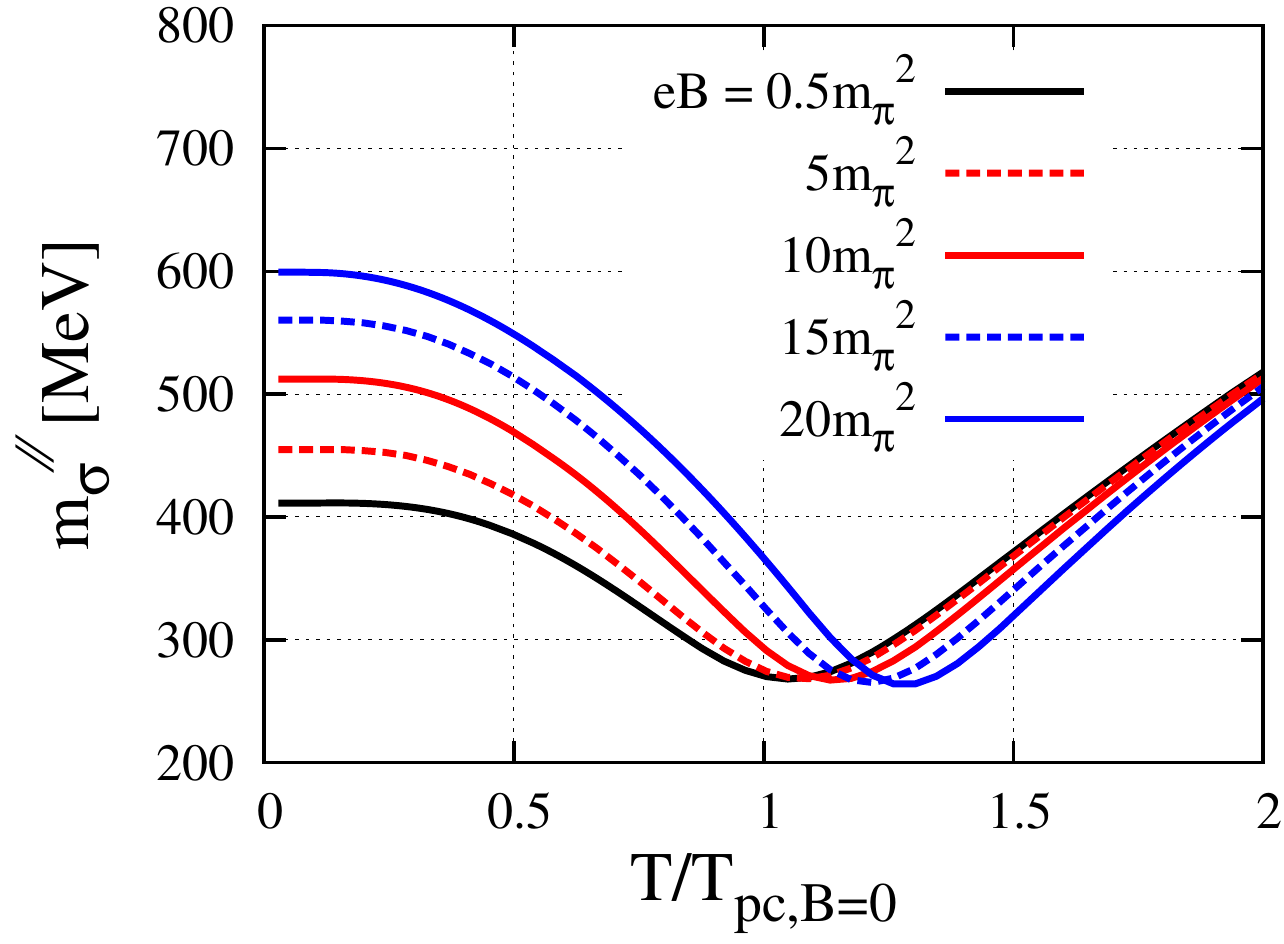}
    \end{center}
   \end{minipage}
   \begin{minipage}{0.5\hsize}
    \begin{center}
     \includegraphics[width=1.00\columnwidth]{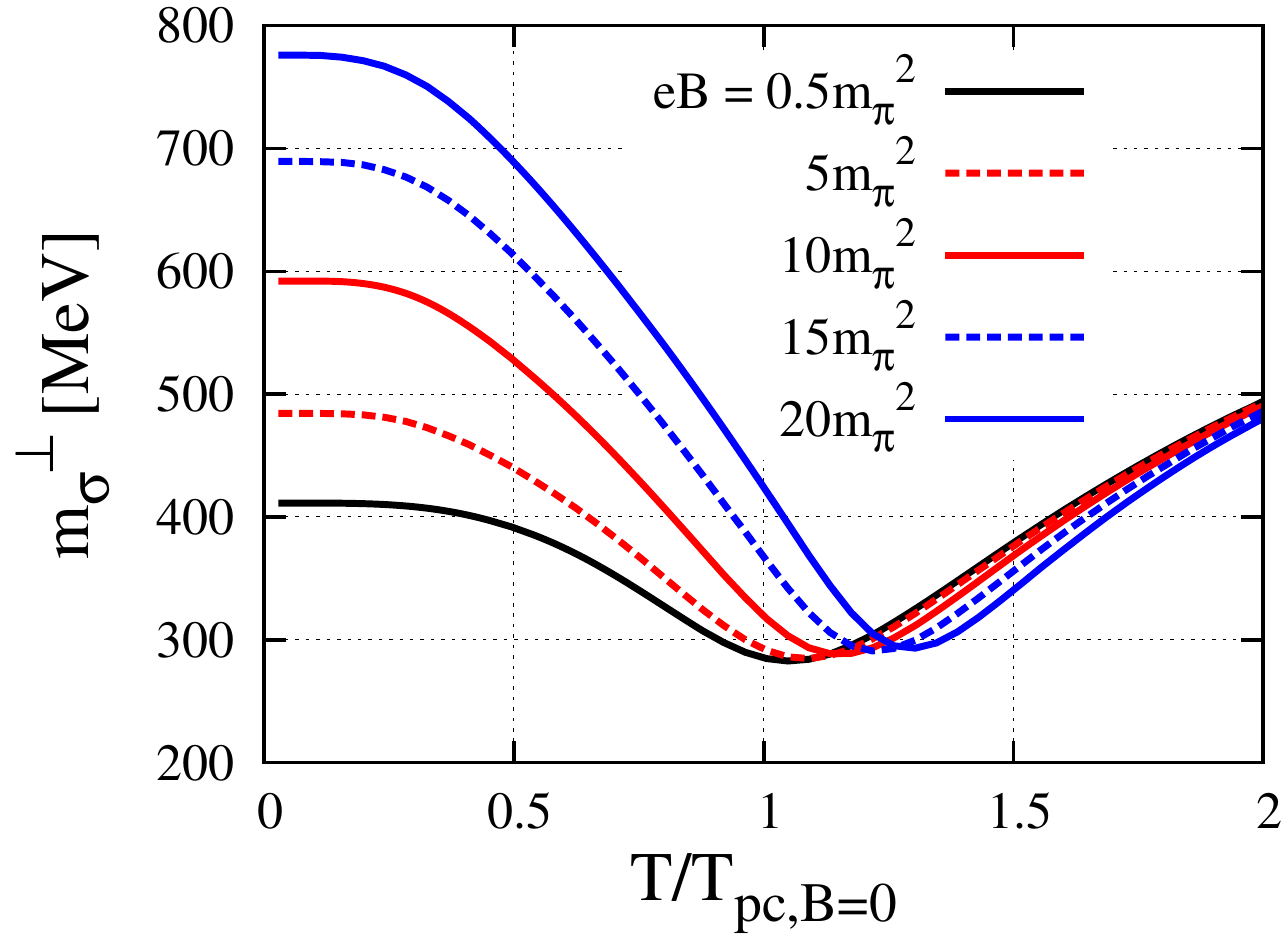}
    \end{center}
   \end{minipage}
   \caption{The longitudinal (left) and transverse (right) sigma
   screening masses with varying external magnetic field.}
   \label{fig:8_m_sigma_vs_T}
  \end{figure}
  Finally, in Fig.~\ref{fig:8_m_sigma_vs_T}, we show the
  direction-dependent renormalized screening masses of the sigma meson.
  Both sigma masses have minimums near the critical temperature.  Above
  $T_{\rm pc}$, the pion and sigma masses for each direction are almost
  degenerate, signaling the effective restoration of chiral symmetry.

  Below $T_{\rm pc}$, $m_{\pi}^{\parallel}$ and $m_{\sigma}^{\perp}$ are
  far more sensitive to the external magnetic field than
  $m_{\pi}^{\perp}$ and $m_{\sigma}^{\parallel}$. The reason is as
  follows. The bare pion mass decreases with the external magnetic field
  while the bare sigma mass increases. On the other hand,
  $Z^{\parallel}$ increases and $Z^{\perp}$ decreases with the external
  magnetic field, respectively. As for $m_{\pi}^{\parallel}$ and
  $m_{\sigma}^{\perp}$, the wave function renormalization and the bare
  meson masses conspire to increase the renormalized masses. Regarding
  $m_{\pi}^{\perp}$ and $m_{\sigma}^{\parallel}$, the effects of the
  wave function renormalization and the bare meson masses interfere with
  each other and the resulting change in the screening mass is reduced.

  Above $T_{\rm pc}$, both the wave function renormalizations and the
  bare meson masses become less sensitive to the external magnetic
  field. Then the renormalized screening masses also become insensitive
  to the external magnetic field.

 \section{Conclusion}
 \label{sc:concl} In the present work, we have examined influences of
 the external magnetic field on the chiral symmetry breaking of strongly
 interacting matter. In order to elucidate the dynamics of neutral
 mesons in the simplest possible setting, we have solved the quark-meson
 model with one light flavor. The quantum and thermal fluctuations of
 mesons and quarks were incorporated with the method of the functional
 renormalization group (FRG) equation.
 
 We have carried out the derivative expansion of the average effective
 action up to second order in the mesonic momentum. With this extended
 truncation, we have successfully taken into account a spatial
 anisotropy of the neutral meson modes which is induced through their
 coupling to quarks. Although this effect has not been considered in
 previous FRG studies
 \cite{Skokov:2011ib,Scherer:2012nn,Fukushima:2012xw,Andersen:2012bq,Andersen:2013swa},
 it is expected to be the origin of the inverse magnetic catalysis
 \cite{Fukushima:2012kc} and our work is the first attempt to test this
 conjecture using FRG.  By devising a novel regulator that is suitable
 for analysis in a magnetic field, we have derived flow equations for
 the scale-dependent effective potential and the wave function
 renormalization at finite temperature and external magnetic field. Then
 we have solved the flow equations numerically using the Taylor
 expansion method, and compared the obtained results with those from the
 leading-order derivative expansion (the so-called LPA) and the
 conventional mean-field approximation.
 
 Our main findings are as follows. 
 \begin{description}\setlength{\itemsep}{-.5mm}
  \item[\quad $\star$] 
	     
 At all temperatures,  the constituent quark mass increases with the
 external magnetic field. Accordingly, the pseudo-critical temperature $T_{\rm pc}$ of chiral restoration 
 is found to increase linearly with the magnetic field. The slope of $T_{\rm pc}$ is close to the 
 mean-field value. We gave a microscopic explanation to this result based on the structure 
 of the flow equations. 
 
 \item[\quad $\star$]
 
 The velocity $v_\perp$ of the neutral mesons moving perpendicular to the magnetic field 
 is found to decrease with the magnetic field at all temperatures, with the largest reduction 
 in $v_\perp$ being observed at zero temperature. In contrast, at high temperature $\gtrsim T_{\rm pc}$, 
 $v_\perp$ becomes rather insensitive to the magnetic field. 
 
  \item[\quad $\star$]
 
 We computed the pion decay constants and the screening
 masses of the neutral mesons for the parallel and perpendicular directions 
 to the external magnetic field. Below $T_{\rm pc}$ they show a 
 large directional dependence, reflecting the anisotropy of the wave function
 renormalizations. 
 \end{description}
 Finally we comment on possible future directions. First and foremost,
 the behavior of $T_{\rm pc}$ in this work is not qualitatively
 consistent with the lattice simulation performed at the physical point
 \cite{Bali:2011qj,Bali:2012zg}, and we must seek for a proper
 explanation of the inverse magnetic catalysis, e.g., in the dynamics of
 gluons which were not taken into account in this work.  Indeed the
 importance of the Polyakov loop was underlined in
 \cite{Bruckmann:2013oba}.  However the preceding analyses
 \cite{Gatto:2010pt,Skokov:2011ib,Andersen:2012jf,Andersen:2013swa} seem
 to suggest that just adding the Polyakov loop in a phenomenological way
 does not resolve the discrepancy with the lattice data. One way to
 address this problem within FRG would be to start from the QCD
 Lagrangian itself rather than effective models.

 It would be also interesting to extend our Ansatz of the effective
 action to two flavors, so that the dynamics of charged mesons is taken
 into account.  From a technical point of view, it is desirable to find
 a more useful regulator function that does not break the rotational
 symmetry explicitly. Finally, to make contact with experiments and
 observations, we should allow for a time-dependent magnetic field and
 evaluate its impact on chiral dynamics.  We leave these issues for
 future work.

 \acknowledgments We are grateful to T. Hatsuda, Y. Hidaka and
 J. Pawlowski for useful discussions. KK was supported by the Special
 Postdoctoral Research Program of RIKEN. TK was supported by RIKEN iTHES
 Project and JSPS KAKENHI Grants Number 25887014.
 
 \appendix
 \section{Derivation of the flow equation for $U_k$}
 \label{sc:der_flow_Uk}

 In this appendix we will give a detailed derivation of
 \eqref{eq:fin_dU}.  First of all, in a purely bosonic constant
 background, the effective action is related to the effective potential
 as $\Gamma_k / V_4 = U_k(\rho) - h\sigma$ where $V_4\equiv \beta L^3$ denotes the
 Euclidean space-time volume.  Consequently, from \eqref{eq:FRG-}, the
 flow equation for the effective potential is obtained as
 \begin{align}
  \label{eq:FRGeq}
  \del_k U_k = \frac{1}{V_4}
  \Bigg\{
  \underbrace{\frac{1}{2}\Tr\left[ \frac{1}{\Gamma^{(2,0)}_k+\RB} 
  \del_k \RB \right]}_{\rm bosons} 
  - \underbrace{\Tr\left[ \frac{1}{\Gamma^{(0,2)}_k+\RF} \del_k \RF
  \right]}_{\rm fermions} \Bigg\}\,.
 \end{align}
 
 \begin{figure}[tb]
  \begin{center}
   \includegraphics[width=0.5\columnwidth]{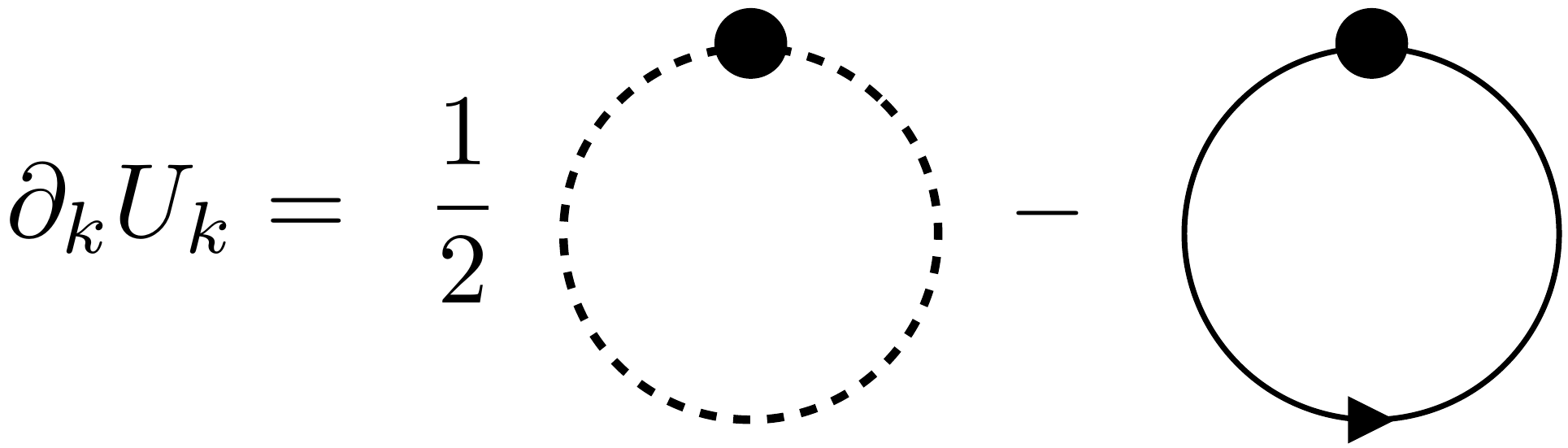}
  \end{center}
  \vspace{-.5\baselineskip}
  \caption{Diagrammatic representation of the flow equation \eqref{eq:FRGeq}. 
  The dashed line (the solid line with arrow) represents  a  
  scale-dependent meson (fermion) propagator, respectively. 
  The black blob stands for the insertion of $\partial_k R_k$.}  
  \label{fig:diagram_frg}
 \end{figure}
 
 The corresponding diagrams are shown in Fig.~\ref{fig:diagram_frg}.
 We note that the dependence of $U_k$ on the magnetic field entirely
 comes from the second term, because the bosons carry no electric
 charge.  The bosonic contribution and the fermionic contribution will
 be evaluated in the appendices \ref{sc:b_T} and \ref{sc:f_T},
 respectively.

  \subsection{Bosonic contribution to $\del_k U_k$}
  \label{sc:b_T}

  From \eqref{eq:FRGeq} and \eqref{eq:RB_3d}, we get
  \begin{align}
   \del_k U_k\Big|_{\rm bose} 
   & = \frac{1}{2}\Tr\left[ \frac{1}{\Gamma^{(2,0)}_k+\RB} \del_k \RB \right]/V_4
   \\
   & = \frac{1}{2}\Tr\left[ \frac{1}{-Z_k^\parallel(\del_4^2+\del_3^2) - 
   Z_k^\perp(\del_1^2+\del_2^2)+\RB+U_k'(\rho)} \del_k \RB \right] /V_4
   \notag
   \\
   & \quad +  
   \frac{1}{2}\Tr\left[ \frac{1}{-Z_k^\parallel(\del_4^2+\del_3^2)-
   Z_k^\perp(\del_1^2+\del_2^2)+\RB+U_k'(\rho)+2\rho U_k''(\rho)} \del_k \RB \right] /V_4
   \\
   & 
   = \frac{1}{2}\ T\!\!\sum_{p_4:\,\text{even}}\int\frac{d^3p}{(2\pi)^3}
   \left\{ (k^2-p_3^2)\del_kZ_k^\parallel + 2kZ_k^\parallel \right\} \theta(k^2-p_3^2) 
   \notag
   \\
   & \quad \times \left[
   \frac{1}{Z_k^\parallel(p_4^2+k^2)+Z_k^\perp p_\perp^2+U_k'(\rho)} + 
   \frac{1}{Z_k^\parallel(p_4^2+k^2)+Z_k^\perp p_\perp^2+U_k'(\rho)+2\rho U_k''(\rho)}
   \right]
   \\
   & = k^2 
   \left(1 + \frac{k}{3}\frac{\del_kZ_k^\parallel}{Z_k^\parallel} \right)
   {\int\,}'\frac{d^2p_\perp}{(2\pi)^3}\left(
   \frac{1}{E_\pi(\rho)}\coth\frac{E_\pi(\rho)}{2T} + 
   \frac{1}{E_\sigma(\rho)}\coth\frac{E_\sigma(\rho)}{2T}
   \right)\,, \hspace{-30pt}
   \label{eq:U_bos}
  \end{align}
  with $p_\perp \equiv (p_1,p_2)$. The definitions of $E_\pi(\rho)$ and
  $E_\sigma(\rho)$ are given in \eqref{eq:EpiEsig}.

  \subsection{Fermionic contribution to $\del_k U_k$}
  \label{sc:f_T}

  From \eqref{eq:FRGeq} and \eqref{eq:RF_3}, we get
  \begin{align}
   \del_k U_k\Big|_{\rm fermi} & 
   = - \Tr\left[ \frac{1}{\Gamma^{(0,2)}_k+\RF} \del_k \RF \right] /V_4
   \\
   & = \scalebox{0.95}{$\displaystyle 
    - N_c \Tr\left[ \frac{1}{\slashed{\del}_4+(\slashed{\del}_3+\RF)+\DD_{\perp}+g(\sigma+i\gamma_5\pi)} \del_k \RF \right] /V_4
   \qquad (\DD_\perp \equiv \gamma_1D_1+\gamma_2D_2)
   $} 
   \\
   & = - N_c \Tr\left[
   \frac{  \slashed{\del}_3+\RF  }
   {\del_4^2+(\slashed{\del}_3+\RF)^2+\DD^2_{\perp}-2g^2\rho} \del_k \RF
   \right]/V_4
   \\
   & = - N_c\ T \!\! \sum_{p_4:\,\text{odd}} \int\frac{dp_3}{2\pi}  \Tr\Bigg[
   \frac{  \displaystyle - i\slashed{p}_3 \frac{k}{|p_3|}  }
   {- p_4^2 -k^2+\DD^2_{\perp}-2g^2\rho} \frac{- i\slashed{p}_3}{|p_3|} 
   \theta(k^2 - p_3^2)
   \Bigg] /L^2
   \\
   & = - \frac{1}{\pi} N_ck^2\ T \!\! \sum_{p_4:\,\text{odd}} \Tr\left[
   \frac{  1  }
   {p_4^2 + k^2 - \DD^2_{\perp} + 2g^2\rho} 
   \right] /L^2  \,.
  \end{align}
  The trace can be evaluated using the eigenfunctions of $\DD^2_{\perp}$, 
  with the result 
  \begin{align}
   \del_k U_k\Big|_{\rm fermi}  
   & = - \frac{2}{\pi}N_ck^2\ T\!\!\sum_{p_4:\,\text{odd}} \Big(\frac{|eB|}{2\pi}\sum_{n=0}^{\infty}\Big)
   \sum_{s=\pm 1/2} \frac{ 1 }{p_4^2 + k^2 + (2n+1-2s)|eB| + 2g^2\rho}
   \label{eq:del_Ufermi}
   \\
   & = - \frac{2}{\pi}N_ck^2  \frac{|eB|}{2\pi}{\sum_{n=0}^{\infty}}'
   \frac{\alpha_n}{2E_n(\rho)}\tanh\frac{E_n(\rho)}{2T} \,,
   \label{eq:del_U_F}
  \end{align}
  with $\alpha_n$ and $E_n(\rho)$ defined in \eqref{eq:alp-E}.  The
  factor 2 in front of \eqref{eq:del_Ufermi} stands for the degeneracy
  of eigenvalues of $\DD^2_{\perp}$ arising from the symmetry
  $[\DD^2_{\perp},\gamma_5] = 0$. As a check, we also computed $\del_k
  U_k\Big|_{\rm fermi}$ using the fermion propagator in a magnetic field
  \eqref{eq:G_ex} and found that the result agrees with
  \eqref{eq:del_U_F} exactly, as it should.

  Finally the sum of \eqref{eq:del_U_F} and \eqref{eq:U_bos} yields
  $\del_k U_k$ in \eqref{eq:fin_dU}.

 \section{Derivation of the flow equations for $Z_k^\perp$ and $Z_k^\parallel$}
 \label{sc:der_flow_Zk}

 \begin{figure}[tb]
  \begin{center}
   \includegraphics[width=0.7\columnwidth]{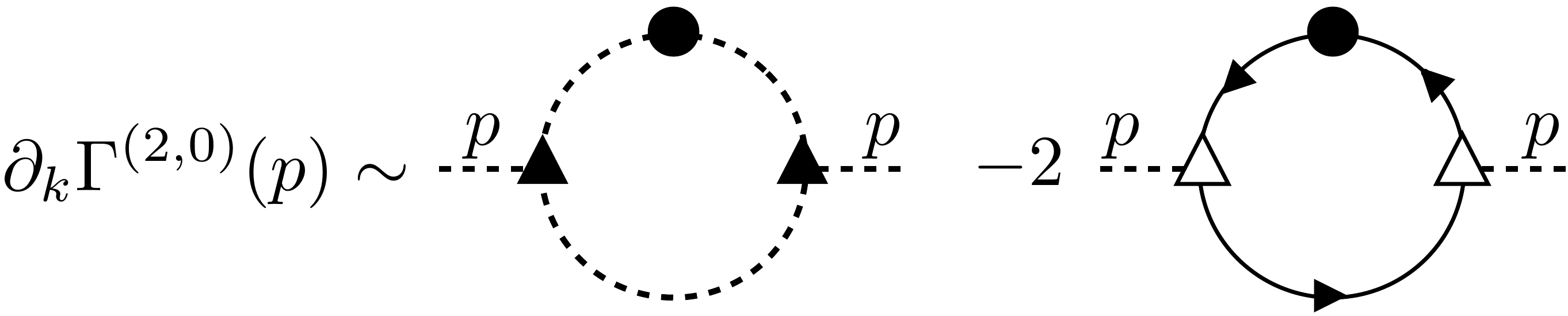}
  \end{center}
  \vspace{-.5\baselineskip}
  \caption{Diagrammatic representation of the flow equation for the
  mesonic two-point function.  The dashed line (the solid line with
  arrow) represents a scale-dependent meson (fermion) propagator,
  respectively.  The black blob stands for the insertion of $\partial_k
  R_k$.  (Another one-loop diagram with a single four-meson vertex is
  not shown here as it does not contribute to the wave function
  renormalization.)  }  \label{fig:dg_frg_2pf}
 \end{figure}

 The flow of $Z_k^\perp$ and $Z_k^\parallel$ receives contribution from
 the diagrams in Fig.~\ref{fig:dg_frg_2pf}. We shall evaluate the
 meson-loop diagram in appendix \ref{sc:bose_Zk} and the fermion-loop
 diagram in appendix \ref{sc:Zfer}.  For brevity we use shorthand
 notations
 \begin{align}
  \int _x \equiv \int_0^{\beta} \! dx_4 \int d^3 x
  \qquad \text{and} \qquad 
  \int_p \equiv T   \sum_{p_4}\int \frac{d^3p}{(2\pi)^3} \,.
 \end{align}

  \subsection{Bosonic contribution to $\del_k Z_k$}
  \label{sc:bose_Zk}

  Let us denote by $\tilde\del_k$ a derivative that only acts on the
  $k$-dependence of the regulator $\RB$. With \eqref{eq:FRG-} and $\RB$
  in \eqref{eq:RB_3d}, the contribution of bosons to the flow equation
  is found to be
  \begin{align}
   \del_k\Gamma_k\Big|_{\rm bose} 
   & = \frac{1}{2} \tilde\del_k \Tr\log[\Gamma_k^{(2,0)}+\RB]
   \\
   & = \frac{1}{2}\tilde\del_k \Tr\log\Bigg[ 
   \underbrace{
   - Z_k^\perp(\del_1^2+\del_2^2) - Z_k^\parallel(\del_3^2+\del_4^2) + \RB
   }_{\equiv\ H_k} +
   \begin{pmatrix} \frac{\del^2 U_k}{\del \sigma^2} & \frac{\del^2
    U_k}{\del \sigma\del\pi} \vspace{2pt}\\ 
   \frac{\del^2 U_k}{\del \pi\del \sigma} & \frac{\del^2 U_k}{\del \pi^2} \end{pmatrix} \Bigg] 
   \\
   & = 
   \frac{1}{2}\tilde\del_k \Tr\log\Bigg[ H_k + 
   \begin{pmatrix}
    U'_k(\rho)+U''_k(\rho)\sigma^2
    & U''_k(\rho)\sigma\pi
   \\ U''_k(\rho)\sigma\pi & U'_k(\rho)+U''_k(\rho)\pi^2 
   \end{pmatrix} \Bigg]\,.
  \end{align}
  We evaluate this in the background $(\sigma,\pi)=(\bar\sigma_k, t(x))$
  where $\bar\sigma_k$ is the running minimum of the potential:
  $\bar\sigma_k\equiv \underset{\sigma}{\mathrm{argmin}}\
  \big\{U_k(\rho)-h\sigma\big\}$\,.  Then
  $\rho=\frac{1}{2}\bar\sigma_k^2+\frac{1}{2}t^2\equiv
  \bar\rho_k+\frac{1}{2}t^2$. Therefore
  \begin{align}
   & \del_k\Gamma_k\Big|_{\rm bose}\Big|_{O(t^2)} 
   \notag
   \\
   = ~ & \frac{1}{2}\tilde\del_k \Tr\log\left[ H_k +  
   \begin{pmatrix} 
    U'_k(\bar\rho_k+\frac{t^2}{2}) + 2\bar\rho_k
    U''_k(\bar\rho_k+\frac{t^2}{2}) &
    U''_k(\bar\rho_k+\frac{t^2}{2})\bar\sigma_k t
    \\ 
    U''_k(\bar\rho_k+\frac{t^2}{2})\bar\sigma_k t &
    U'_k(\bar\rho_k+\frac{t^2}{2})+U''_k(\bar\rho_k+\frac{t^2}{2}) t^2 
   \end{pmatrix}
   \right] \Bigg|_{O(t^2)}
   \\
   = ~ & \frac{1}{2}\tilde\del_k \Tr\log\left[ A+B+C \right]\Big|_{O(t^2)}
   \\
   = ~ & \frac{1}{2}\tilde\del_k \Tr \left[A^{-1}C - \frac{1}{2} A^{-1} B A^{-1}B \right] \,,
   \label{eq:acab}
  \end{align}
  with the definitions
  \begin{align}
	      A & \equiv \begin{pmatrix} H_k + U'_k(\bar\rho_k)+2\bar\rho_k U''_k(\bar\rho_k) & 0 \\ 0 & H_k + U'_k(\bar\rho_k) \end{pmatrix}
	  \equiv \begin{pmatrix} H_k + \zetax & 0 \\ 0 & H_k + \zetay \end{pmatrix}
   \,,
   \\
	      B & \equiv \begin{pmatrix} 0 &
			 U''_k(\bar\rho_k)\bar\sigma_k t \\
			 U''_k(\bar\rho_k)\bar\sigma_k t & 0
			 \end{pmatrix}\,,
   \\
   C & \equiv \begin{pmatrix} \frac{1}{2}U''_k(\bar\rho_k)t^2 
	      + \bar\rho_k U'''_k(\bar\rho_k)t^2 & 0 \\ 0 & \frac{3}{2}U''_k(\bar\rho_k) t^2  \end{pmatrix}\,.
  \end{align}
  The first term in \eqref{eq:acab} can be neglected as it does not
  generate a kinetic term $\sim t_pt_{-p}$. As for the second term,
  \begin{align}
   \frac{1}{2}\tilde\del_k \Tr \left[- \frac{1}{2} A^{-1} B A^{-1}B \right] 
   & = - \bar\rho_k\big[U''_k(\bar\rho_k)\big]^2\tilde\del_k \Tr \left[A_{11}^{-1}tA_{22}^{-1}t\right]
   \\
   & = - \bar\rho_k\big[U''_k(\bar\rho_k)\big]^2\tilde\del_k \int_{pq} 
   (A_{11}^{-1})_p t_{p-q} (A_{22}^{-1})_q t_{q-p}
   \\
   & = - \bar\rho_k\big[U''_k(\bar\rho_k)\big]^2\tilde\del_k \int_{p} t_p t_{-p} \int_q  
   (A_{11}^{-1})_{p+q} (A_{22}^{-1})_q  \,.
   \label{eq:truei}
  \end{align}
  On the other hand, we have
  \begin{align}
   \del_k\Gamma_k 
   = ~ & \del_k \int_x \Big\{
   \frac{Z_k^\perp}{2} \sum_{i=1,2} (\del_i t)^2  
   + \frac{Z_k^\parallel }{2} \sum_{i=3,4} (\del_i t)^2 + \dots 
   \Big\}
   \\
   =~ & \frac{1}{2}\del_k Z_k^\perp  \int_p t_p t_{-p} (p_1^2+p_2^2) 
   + \frac{1}{2}\del_k Z_k^\parallel  \int_p t_p t_{-p} (p_3^2+p_4^2)
   + \dots \,.
   \label{eq:tt}
  \end{align}  
  Comparing \eqref{eq:tt} with \eqref{eq:truei}, we are led to the important formulae
  \begin{align}
   \del_k Z_k^\perp \Big|_{\rm bose} & = 
   -  \bar\rho_k\big[U''_k(\bar\rho_k)\big]^2\lim_{p\to 0}\frac{\del^2}{\del p_1^2}
   \tilde\del_k \int_q (A_{11}^{-1})_{p+q} (A_{22}^{-1})_q \,,
   \label{eq:116+-}
   \\
   \del_k Z_k^\parallel \Big|_{\rm bose} & = 
   -  \bar\rho_k\big[U''_k(\bar\rho_k)\big]^2\lim_{p\to 0}\frac{\del^2}{\del p_3^2}
   \tilde\del_k \int_q (A_{11}^{-1})_{p+q} (A_{22}^{-1})_q \,. 
   \label{eq:116+}
  \end{align}
  Without loss of generality one can assume $p=(p_1,0,p_3,0)$. With a bit of algebra, we find  
  \begin{align}
   & \tilde\del_k  \int_q  (A_{11}^{-1})_{\pp} (A_{22}^{-1})_q \Big|_{\pp=q+p}
   \notag
   \\
   = ~ & \scalebox{0.9}{$\displaystyle \tilde\del_k \int_q 
   \frac{1}{Z_k^\perp(\pp_1^2+\pp_2^2)+Z_k^\parallel(\pp_3^2+\pp_4^2)+\RB(\pp) + \zetax } ~
   \frac{1}{Z_k^\perp(q_1^2+q_2^2)+Z_k^\parallel(q_3^2+q_4^2)+\RB(q) + \zetay} \Bigg|_{\pp=q+p}
   $}
   \\
   = ~ & - T\sum_{q_4:\,\text{even}} \Big\{ I(Z_k^\parallel q_4^2+\zetax, Z_k^\parallel q_4^2+\zetay; p) + 
   (\zetax \leftrightarrow \zetay) \Big\} \,, 
   \label{eq:121-}
  \end{align}
  with 
  \begin{align}
   &\hspace{-20pt} I(\zeta,\zeta';p) \equiv \notag
   \\
   & \hspace{-20pt}\scalebox{0.9}{$\displaystyle  \int \frac{d^3q}{(2\pi)^3}
   \frac{\big[ 2kZ_k^\parallel + (k^2-q_3^2)\del_kZ_k^\parallel \big]\,\theta(k^2-q_3^2)}
   {\big[ Z_k^\perp(q_1^2+q_2^2) + Z_k^\parallel k^2 + \zeta \big]^2 \Big\{
   Z_k^\perp\big((q_1-p_1)^2+q_2^2\big) + Z_k^\parallel (q_3-p_3)^2 +\RB(q-p) + \zeta'
   \Big\}  } 
   \,. $}\hspace{-20pt}
  \end{align}
  The next task is to extract the $O(p^2)$ part of $I(\zeta,\zeta';p)$. To take care of  
  $\RB(q-p)$ in the denominator, we decompose this integral into two pieces as 
  $I(\zeta,\zeta';p)=I_1(\zeta,\zeta';p)+I_2(\zeta,\zeta';p)$, with 
  \begin{align}
   & I_1(\zeta,\zeta';p) \equiv  \notag
   \\
   & ~~ \scalebox{0.85}{$\displaystyle  \int \frac{d^3q}{(2\pi)^3}
   \frac{\big[ 2kZ_k^\parallel + (k^2-q_3^2)\del_kZ_k^\parallel \big]\,\theta(k^2-q_3^2)~{\color{blue}\theta\big(k^2-(q_3-p_3)^2\big)}}
   {\big[ Z_k^\perp(q_1^2+q_2^2) + Z_k^\parallel k^2 + \zeta \big]^2 \Big\{
   Z_k^\perp\big((q_1-p_1)^2+q_2^2\big) + Z_k^\parallel (q_3-p_3)^2 +\RB(q-p) + \zeta' \Big\}  }\,,
   $}
   \\
   & I_2(\zeta,\zeta';p) \equiv \notag
   \\ 
   & ~~ \scalebox{0.85}{$\displaystyle \int \frac{d^3q}{(2\pi)^3}
   \frac{\big[ 2kZ_k^\parallel + (k^2-q_3^2)\del_kZ_k^\parallel
   \big]\,\theta(k^2-q_3^2)~{\color{blue} \big\{1-\theta\big(k^2-(q_3-p_3)^2\big)\big\}}}
   {\big[ Z_k^\perp(q_1^2+q_2^2) + Z_k^\parallel k^2 + \zeta \big]^2 \Big\{
   Z_k^\perp\big((q_1-p_1)^2+q_2^2\big) + Z_k^\parallel (q_3-p_3)^2 +\RB(q-p) + \zeta'  \Big\}  }\,,
   $} 
  \end{align}
  A straightforward but tedious calculation yields  
  \begin{align}
   & \lim_{p\to 0}\frac{\del^2}{\del p_1^2}I_1(\zeta,\zeta';p) = 
   \notag 
   \\
   & ~ - 16 k^2 \big(Z_k^\perp\big)^2 
   \left(Z_k^\parallel + \frac{k}{3} \del_kZ_k^\parallel \right)  
   \int \frac{d^2q_\perp}{(2\pi)^3}
   \frac{q_\perp^2}{( Z_k^\perp q_\perp^2 + Z_k^\parallel k^2 + \zeta )^3
   (Z_k^\perp q_\perp^2 + Z_k^\parallel k^2 + \zeta' )^2}\,, 
   \hspace{-20pt}
   \\
   & \lim_{p\to 0}\frac{\del^2}{\del p_3^2} I_1(\zeta,\zeta';p) 
   = -2k \del_k Z_k^\parallel 
   \int \frac{d^2q_\perp}{(2\pi)^3}
   \frac{1}{( Z_k^\perp q_\perp^2 + Z_k^\parallel k^2 + \zeta )^2 
   ( Z_k^\perp q_\perp^2 + Z_k^\parallel k^2 + \zeta' )  }\,, 
   \label{eq:I1p3__}
   \\
   & \lim_{p\to 0}\frac{\del^2}{\del p_1^2} I_2(\zeta,\zeta';p) = 0\,,
   \\
   & \lim_{p\to 0}\frac{\del^2}{\del p_3^2} I_2(\zeta,\zeta';p) = 
   \notag
   \\
   & ~~\int \frac{d^2q_\perp}{(2\pi)^3}
   \frac{1}{( Z_k^\perp q_\perp^2 + Z_k^\parallel k^2 + \zeta )^2} \left(
   -\frac{4k^2 (Z_k^\parallel)^2}{(Z_k^\perp q_\perp^2+Z_k^\parallel k^2+\zeta')^2}
   + \frac{2k\del_k Z_k^\parallel}{Z_k^\perp q_\perp^2 + Z_k^\parallel k^2 + \zeta'}
   \right)\,. 
   \label{eq:I2p3-}
  \end{align}
  Combining all the above and performing a change of variable
  ($q_\perp^2 \to \frac{Z_k^\parallel}{Z_k^\perp}q_\perp^2$), we get
  \begin{align}
   \lim_{p\to 0}\frac{\del^2}{\del p_1^2}I(\zeta,\zeta';p) & = 
   - 16 \frac{k^2}{(Z_k^\parallel)^2}  \left(1 + \frac{k}{3} \frac{\del_kZ_k^\parallel}{Z_k^\parallel} \right)  
   \int \frac{d^2q_\perp}{(2\pi)^3}
   \frac{q_\perp^2}{\Big( q_\perp^2 + k^2 + \frac{\zeta}{Z_k^\parallel} \Big)^3
   \Big( q_\perp^2 + k^2 + \frac{\zeta'}{Z_k^\parallel} \Big)^2}
   \\
   & = - \frac{2}{\pi^2}\frac{k^2}{(Z_k^\parallel)^2}  \left(1 + \frac{k}{3} \frac{\del_kZ_k^\parallel}{Z_k^\parallel} \right)
   \int_0^\infty \!\! dw \frac{w}{\Big( w + k^2 + \frac{\zeta}{Z_k^\parallel} \Big)^3
   \Big( w + k^2 + \frac{\zeta'}{Z_k^\parallel} \Big)^2}
   \,,
   \\
   \lim_{p\to 0}\frac{\del^2}{\del p_3^2}I(\zeta,\zeta';p) & = 
   - 4 \frac{k^2}{Z_k^\parallel Z_k^\perp} \int \frac{d^2q_\perp}{(2\pi)^3}
   \frac{1}{\Big( q_\perp^2 + k^2 + \frac{\zeta}{Z_k^\parallel} \Big)^2
   \Big( q_\perp^2 + k^2 + \frac{\zeta'}{Z_k^\parallel}\Big)^2} 
   \\
   & = - \frac{1}{2\pi^2}\frac{k^2}{Z_k^\parallel Z_k^\perp} \int_0^\infty \!\! 
   \frac{dw}{\Big( w + k^2 + \frac{\zeta}{Z_k^\parallel} \Big)^2
   \Big( w + k^2 + \frac{\zeta'}{Z_k^\parallel}\Big)^2}
   \,.
  \end{align}
  Plugging these into \eqref{eq:116+-} and \eqref{eq:116+}, we find 
  \begin{align}
   & \del_k Z_k^\perp \Big|_{\rm bose} 
   \notag 
   \\
   = & ~ \bar\rho_k\big[U''_k(\bar\rho_k)\big]^2\, T\!\! \sum_{q_4:\,\text{even}}
   \lim_{p\to 0}\frac{\del^2}{\del p_1^2}\Big\{ I(Z_k^\parallel
   q_4^2+\zetax, Z_k^\parallel q_4^2+\zetay;p) +
   (\zetax \leftrightarrow \zetay) \Big\}
   \\
   = & - \frac{k^2}{\pi^2} \frac{\bar\rho_k\big[U''_k(\bar\rho_k)\big]^2}{(Z_k^\parallel)^2}
   \left(1 + \frac{k}{3} \frac{\del_kZ_k^\parallel}{Z_k^\parallel} \right) T\!\! \sum_{q_4:\,\text{even}} 
   \int_0^\infty \!\! 
   \frac{dw}{\Big( w + k^2 + q_4^2 + \frac{\zetax}{Z_k^\parallel} \Big)^2
   \Big( w + k^2 + q_4^2 + \frac{\zetay}{Z_k^\parallel}\Big)^2}\,, 
   \label{eq:Zpp}
\end{align}
and
\begin{align}
   & \del_k Z_k^\parallel \Big|_{\rm bose} 
   \notag
   \\
   = &~ \bar\rho_k\big[U''_k(\bar\rho_k)\big]^2 \,T\!\!\sum_{q_4:\,\text{even}}
   \lim_{p\to 0}\frac{\del^2}{\del p_3^2}\Big\{ I(Z_k^\parallel q_4^2+\zetax, Z_k^\parallel q_4^2+\zetay;p) + 
   (\zetax \leftrightarrow \zetay ) \Big\}  
   \\
   = & - \frac{k^2}{\pi^2}\frac{\bar\rho_k\big[U''_k(\bar\rho_k)\big]^2}{Z_k^\parallel Z_k^\perp}  
   \,T \!\! \sum_{q_4:\,\text{even}} \int_0^\infty \!\! 
   \frac{dw}{\Big( w + k^2 + q_4^2 + \frac{\zetax}{Z_k^\parallel} \Big)^2
   \Big( w + k^2 + q_4^2 + \frac{\zetay}{Z_k^\parallel}\Big)^2}\,.
   \label{eq:dZ_b2}
  \end{align}
  In deriving \eqref{eq:Zpp} we have used a mathematical formula 
  \begin{align}
   \int_0^\infty \!\! dw \frac{w}{(w+\alpha)^3(w+\beta)^2} + \int_0^\infty \!\! dw \frac{w}{(w+\alpha)^2(w+\beta)^3} 
   = \frac{1}{2}\int_0^\infty \!\! \frac{dw}{(w+\alpha)^2(w+\beta)^2} 
  \end{align}
  which holds for $\alpha>0$ and $\beta>0$. 

  The expressions \eqref{eq:Zpp} and \eqref{eq:dZ_b2} are not so useful
  for numerical analysis since they involve infinite sums as well as
  integrals over the whole real axis. One can simplify them by using
  Feynman's integral formula and then taking the Matsubara sums
  analytically:
  \begin{align}
   & T \!\! \sum_{q_4:\,\text{even}} \int_0^\infty \!\! 
   \frac{dw}{\Big( w + k^2 + q_4^2 + \frac{\zetax}{Z_k^\parallel} \Big)^2
   \Big( w + k^2 + q_4^2 + \frac{\zetay}{Z_k^\parallel}\Big)^2}
   \notag
   \\
   = ~ & T \!\! \sum_{q_4:\,\text{even}} \int_0^1dx ~ 6x(1-x) 
   \int_0^\infty \!\! \frac{dw}{
   \Big(w+k^2+q_4^2+\frac{x\zetax+(1-x)\zetay}{Z_k^\parallel} \Big)^4
   }
   \\
   = ~ & T \!\! \sum_{q_4:\,\text{even}} \int_0^1dx ~ 2x(1-x) \frac{1}{
   \Big(k^2+q_4^2+\frac{x\zetax+(1-x)\zetay}{Z_k^\parallel} \Big)^3  }
   \\
   = ~ & \int_0^1dx ~ 2x(1-x) 
   \frac{\displaystyle 
   \left( 6T^2+Q^2 \mathrm{csch}^2\frac{Q}{2T} \right)
   \mathrm{coth}\frac{Q}{2T} + 3TQ \cdot \mathrm{csch}^2\frac{Q}{2T}
   }{32\cdot T^2 \cdot Q^5} \,,
  \end{align}
  with 
  \begin{align}
   Q = Q(x) \equiv \sqrt{\displaystyle  k^2+\frac{x \zetax + (1-x)\zetay}{Z_k^\parallel}} \,.
  \end{align}
  We used this expression in our actual numerical computation.

  \subsection{Fermionic contribution to $\del_k Z_k$}
  \label{sc:Zfer}

  From \eqref{eq:FRG-}, 
  \begin{align}
   \del_k\Gamma_k\Big|_{\rm fermi} & = - \tilde\del_k \Tr\log[\Gamma_k^{(0,2)}+\RF]
   \\
   & = - N_c \, \tilde\del_k \Tr\log[ \DD + \RF + g (\sigma + i\gamma_5 \pi)]\,.
  \end{align}
  As in the previous section, we evaluate this in an inhomogeneous background $(\sigma(x),\pi(x)) 
  = (\bar\sigma_k, t(x))$, with $\bar\sigma_k$ the running minimum of the potential: 
  $\bar\sigma_k\equiv \underset{\sigma}{\mathrm{argmin}}\ \big\{U_k(\rho)-h\sigma\big\}$\,. Then
  \begin{align}
   \del_k\Gamma_k\Big|_{\rm fermi} 
   & = -  N_c\, \tilde\del_k \Tr\log[ \DD + \RF + g\bar\sigma_k + ig \gamma_5 t] \,. 
   \label{eq:3f4r}
  \end{align}
  Introducing the regulator-dependent fermion propagator 
  $\displaystyle G\equiv \frac{1}{\DD +\RF + g\bar\sigma_k}$ one can expand \eqref{eq:3f4r} 
  to $O(t^2)$ to obtain
  \begin{align}
   \del_k\Gamma_k\Big|_{\rm fermi}\Big|_{O(t^2)} 
   & =  - \frac{1}{2}N_c g^2\tilde \del_k  \Tr[G \gamma_5 t G \gamma_5 t] 
   \label{eq:G_ss}
   \\
   & = - \frac{1}{2}N_c g^2\tilde \del_k  \Tr[\tilde G \gamma_5 t \tilde G \gamma_5 t] 
   \\
   & = - \frac{1}{2}N_c g^2 \int_{p} 
   t_{p}t_{-p} ~ \tilde \del_k \int_{q} \tr [\tilde G_{p+q}\gamma_5 \tilde G_q \gamma_5 ]\,, 
   \label{eq:G_343}
  \end{align}
  where `$\tr$' is a trace over spinor indices, and $\tilde G$ is
  the \emph{translationally invariant part} of $G$. This replacement is
  justified because the so-called Schwinger phase \cite{Schwinger:1951nm} in $G$
  drops out of the trace in \eqref{eq:G_ss}.

  Comparing \eqref{eq:G_343} with \eqref{eq:tt} we obtain 
  \begin{align}
   \label{eq:Zflow}
   \del_k Z_k^\perp \Big|_{\rm fermi} & = - N_c g^2 \int_{q} 
   \tilde \del_k \lim_{p\to 0}\frac{d}{d p_1^2} \tr [\tilde G_{p+q}\gamma_5 \tilde G_q \gamma_5 ]\,,
   \\
   \label{eq:Zflow2}
   \del_k Z_k^\parallel \Big|_{\rm fermi} & = - N_c g^2 \int_{q} 
   \tilde \del_k \lim_{p\to 0}\frac{d}{d p_3^2} \tr [\tilde
   G_{p+q}\gamma_5 \tilde G_q \gamma_5 ]\,.
  \end{align}
  A closed expression for $\tilde G_p$ was derived in
  \cite{Schwinger:1951nm} in the absence of the regulator $\RF$ (see
  also \cite{Gusynin:1995nb,Shovkovy:2012zn}). The formula, after
  analytic continuation to the Euclidean space-time, reads
  \begin{align}
   \tilde G_p\Big|_{\RF=0} & = \frac{1}{\DD+g\bar\sigma_k} (p)
   \\
   & = 2 \exp\left(-\frac{p_\perp^2}{|eB|}\right) \sum_{n=0}^\infty \frac{(-1)^n 
   D_n(p)}{p_4^2 + p_3^2+ 2|eB|n  + g^2\bar\sigma_k^2}\,,
  \end{align}
  with $p_\perp^2 \equiv p_1^2+p_2^2$ and 
  \begin{align} 
   D_n(p) & \equiv \big[ i(p_4 \gamma_4+p_3\gamma_3)+g\bar\sigma_k \big] 
   \left\{  L_n\left(\frac{2p_\perp^2}{|eB|}\right)\mathcal{P}_+ - 
   L_{n-1}\left(\frac{2p_\perp^2}{|eB|}\right)\mathcal{P}_-  \right\}  
   \notag
   \\
   & \quad - 2i (p_1\gamma_1+p_2\gamma_2)L_{n-1}^{(1)}\left(\frac{2p_\perp^2}{|eB|}\right)\,. 
  \end{align}
  Here $\mathcal{P}_\pm\equiv \frac{1}{2}(\1\mp i\gamma_1\gamma_2)$ are
  the spin projectors%
   \footnote{The relative sign is reversed owing to
  the Euclidean convention. In this work we are using Hermitian gamma
  matrices defined as $\{\gamma_\mu, \gamma_\nu\}=2\delta_{\mu\nu}$ with
  $(\gamma_\mu)^\dagger = \gamma_\mu$.} while $L_n(x)$ and
  $L_n^{(\alpha)}(x)$ are the (generalized) Laguerre polynomials. In
  what follows, we promise $L_n (x) = L_n^{(\alpha)}(x) \equiv 0$ for
  $n<0$.

  In the presence of the regulator $\RF$ in \eqref{eq:RF_3} the
  propagator is modified as $(\DD+g\bar\sigma_k)^{-1} \to
  (\DD+\RF+g\bar\sigma_k)^{-1}=\big(-i\slashed{p}_3[1+r_k(p_3)]+\dots\big)^{-1}$.
  It follows that one can incorporate $\RF$ into the propagator by
  simply replacing $p_3$ with $p_3[1+r_k(p_3)]$. Therefore we have
  \begin{align}
   \tilde G_p & = 2 \exp\left(-\frac{p_\perp^2}{|eB|}\right) \sum_{n=0}^\infty 
   \frac{(-1)^n D_n^{(k)}(p)}{p_4^2 + p_3^2[1+r_k(p_3)]^2 + 2|eB|n + g^2\bar\sigma_k^2}\,,
   \label{eq:G_ex}
  \end{align}
  with $D_n^{(k)}(p) \equiv D_n\big(p_1, p_2, p_3[1+r_k(p_3)],p_4\big)$\,.  

  Before proceeding, let us introduce shorthand notations for some
  useful quantities:
  \begin{align}
   \qq_3 & \equiv q_3+p_3\,, 
   \\
   F_n(q_4,q_3) & \equiv q_4^2 + q_3^2[1+r_k(q_3)]^2 + 2|eB|n + g^2\bar\sigma_k^2 \,, 
   \label{eq:F_def}
   \\
   \FF_0(q_4,q_3,\qq_3) & \equiv 
   q_4^2+\qq_3 q_3[1+r_k(\qq_3)][1+r_k(q_3)] + g^2\bar\sigma_k^2 \,. 
  \end{align}
  Our remaining task is to plug \eqref{eq:G_ex} into \eqref{eq:Zflow}
  and \eqref{eq:Zflow2}.  As this is a lengthy calculation we divide
  this into a few smaller steps. Firstly, we have from \eqref{eq:G_ex}
  \begin{align} 
   \tr [\tilde G_{p+q}\gamma_5 \tilde G_q \gamma_5 ] 
   = ~ & 4 \exp\left(-\frac{(q+p)_\perp^2+q_\perp^2}{|eB|}\right) \sum_{m=0}^\infty \sum_{n=0}^\infty 
   \frac{
   (-1)^{m+n} \tr \big[ D_m^{(k)}(q+p) \gamma_5 D_n^{(k)}(q) \gamma_5 \big]
   }{  F_m(q_4,\qq_3)F_n(q_4,q_3)  }  \,. 
   \label{eq:98012}
  \end{align} 
  Without loss of generality we may assume $p=(p_1,0,p_3,0)$.  
  With a bit of algebra, the trace in the numerator becomes
  \begin{multline} 
   \tr \big[ D_m^{(k)}(q+p) \gamma_5 D_n^{(k)}(q) \gamma_5 \big]
   \\
   =  
   2 \FF_0(q_4,q_3,\qq_3)
   \Bigg\{
   L_m\left(\frac{2(q+p)_\perp^2}{|eB|}\right)
   L_n\left(\frac{2q_\perp^2}{|eB|}\right) 
   + 
   L_{m-1}\left(\frac{2(q+p)_\perp^2}{|eB|}\right)
   L_{n-1}\left(\frac{2q_\perp^2}{|eB|}\right)  
   \Bigg\}
   \\
   +16 (q_\perp^2+q_1p_1) L_{m-1}^{(1)}\left(\frac{2(q+p)_\perp^2}{|eB|} \right)
   L_{n-1}^{(1)}\left(\frac{2q_\perp^2}{|eB|}\right)
   \,,
  \end{multline}
  and the integration over the transverse momenta yields
  \begin{align}
   & \int \frac{d^2q_\perp}{(2\pi)^2} 
   \exp \left(-\frac{(q+p)_\perp^2+q_\perp^2}{|eB|}\right)
   \tr \big[ D_m^{(k)}(q+p) \gamma_5 D_n^{(k)}(q) \gamma_5 \big]
   \notag
   \\
   & =  \FF_0(q_4,q_3,\qq_3) 
   \frac{|eB|}{4\pi}\ee^{-W}(-W)^{m-n} 
   \times 
   \notag
   \\
   & \qquad \times \left\{
   \frac{n!}{m!}\left[L_n^{(m-n)}(W)\right]^2\theta(m,n\geq 0) + 
   \frac{(n-1)!}{(m-1)!}\left[L_{n-1}^{(m-n)}(W)\right]^2\theta(m,n \geq 1)
   \right\}
   \notag
   \\
   & \quad + \frac{|eB|^2}{\pi}\ee^{-W}(-W)^{m-n}\frac{n!}{(m-1)!} L_{n-1}^{(m-n)}(W) L_n^{(m-n)}(W)
   ~\theta(m,n\geq 1) 
   \label{eq:54324}
  \end{align}
  where $\displaystyle W \equiv \frac{p_1^2}{2|eB|}$, and
  $\theta(\bullet)$ is defined as unity if $(\bullet)$ is true, and
  zero otherwise. In deriving \eqref{eq:54324} we have used two
  mathematical formulas:
  \begin{align}
   & \int \frac{d^2q_\perp}{(2\pi)^2} \exp \left(-\frac{(q+p)_\perp^2+q_\perp^2}{|eB|}\right)
   L_k\left(\frac{2q_\perp^2}{|eB|}\right)L_\ell\left(\frac{2(q+p)_\perp^2}{|eB|}\right)
   \notag
   \\
   & =   
   \frac{|eB|}{8\pi}\frac{k!}{\ell !}\ee^{-
   \frac{p_\perp^2}{2|eB|}}\left(- \frac{p_\perp^2}{2|eB|} \right)^{\ell
   - k}\left[ L_k^{(\ell-k)}\left( \frac{p_\perp^2}{2|eB|} \right)
   \right]^2
   \qquad\qquad  \text{for}~~k,\,\ell\geq 0\,,
  \end{align}
  and
  \begin{align}
   & \int \frac{d^2q_\perp}{(2\pi)^2} \exp \left(-\frac{(q+p)_\perp^2+q_\perp^2}{|eB|}\right)
   (q_\perp^2+q_\perp\cdot p_\perp)
   L^{(1)}_{k-1}\left(\frac{2q_\perp^2}{|eB|}\right)L^{(1)}_{\ell-1}\left(\frac{2(q+p)_\perp^2}{|eB|}\right)
   \notag
   \\
   & =   
   \frac{|eB|^2}{16\pi}\frac{k!}{\ell !}\ee^{- \frac{p_\perp^2}{2|eB|} }
   \left(- \frac{p_\perp^2}{2|eB|} \right)^{\ell - k} \ell 
   L_{k-1}^{(\ell-k)}\left( \frac{p_\perp^2}{2|eB|} \right) 
   L_k^{(\ell-k)}\left( \frac{p_\perp^2}{2|eB|} \right) 
   \qquad \qquad \text{for}~~k,\,\ell\geq 1\,. 
  \end{align}
  From \eqref{eq:98012} and \eqref{eq:54324} we obtain
  \begin{align}
   & \int \frac{d^2q_\perp}{(2\pi)^2} \tr [\tilde G_{p+q}\gamma_5 \tilde G_q \gamma_5 ]
   \notag
   \\
   & = 
   \frac{|eB|}{\pi} \FF_0(q_4,q_3,\qq_3)\sum_{m=0}^{\infty} \sum_{n=0}^{\infty}
   \frac{\ee^{-W}W^{m-n}}{F_m(q_4,\qq_3)F_n(q_4,q_3)}
   \times 
   \notag
   \\
   & \qquad \qquad  \times \left\{
   \frac{n!}{m!}\left[L_n^{(m-n)}(W)\right]^2  + 
   \frac{(n-1)!}{(m-1)!}\left[L_{n-1}^{(m-n)}(W)\right]^2\theta(m,n \geq 1)
   \right\}
   \notag
   \\
   & \quad + 
   \frac{4}{\pi} |eB|^2
   \sum_{m=1}^{\infty} \sum_{n=1}^{\infty}
   \frac{\ee^{-W}W^{m-n}}{F_m(q_4,\qq_3)F_n(q_4,q_3)}
   \frac{n!}{(m-1)!} L_{n-1}^{(m-n)}(W) L_n^{(m-n)}(W)\,.
   \label{eq:113-}
  \end{align}
  From here on we shall consider $\del_kZ_k^\perp\Big|_{\rm fermi}$
  and $\del_kZ_k^\parallel\Big|_{\rm fermi}$ separately.

   \subsubsection{Flow of $Z_k^\perp$}
   \label{sc:z_perp_f}

   In the following we set $p_3=0$ without losing generality. Let us
   rewrite \eqref{eq:Zflow} as
   \begin{align}
    \hspace{-20pt}
    \del_k Z_k^\perp \Big|_{\rm fermi} & = - N_c g^2 \, 
    T\!\! \sum_{q_4:\,\text{odd}} \int \frac{dq_3}{2\pi} 
    \ \tilde \del_k \  \lim_{p\to 0}\frac{d}{d p_1^2} 
    \int \frac{d^2q_\perp}{(2\pi)^2} 
    \tr [\tilde G_{p+q}\gamma_5 \tilde G_q \gamma_5 ]
    \\
    & = - N_c g^2 \, 
    T\!\! \sum_{q_4:\,\text{odd}} \int \frac{dq_3}{2\pi} 
    \ \tilde \del_k\left( \frac{1}{2|eB|} \lim_{W\to 0}\frac{d}{d W} 
    \int \frac{d^2q_\perp}{(2\pi)^2} 
    \tr [\tilde G_{p+q}\gamma_5 \tilde G_q \gamma_5 ] \right) \,.
    \label{eq:ff34534}
    \hspace{-20pt}
   \end{align}
   We will carry out these operations in this order. First,
   recall that for $\alpha\in\mathbb{Z}$,
   \begin{align}
    L_n^{(\alpha)}(x) ~\underset{x\to 0}{\sim}~
    \left\{
    \begin{array}{cl}
     \begin{pmatrix}n+\alpha\\n\end{pmatrix} & [\alpha\geq 0]
     \\
     \displaystyle 
      \frac{(-1)^{\alpha}}{(-\alpha)!}x^{-\alpha} &  [-n\leq \alpha \leq -1]
      \\
	    (-1)^n \begin{pmatrix}-\alpha-1\\n\end{pmatrix} & [\alpha\leq -n-1]
    \end{array}
    \right..
    \label{eq:Lagu}
   \end{align}
   Using this, it is not difficult to show that all terms in
   \eqref{eq:113-} with $|m-n|\geq 2$ are $O(W^2)$ in the limit $W\to
   0$ and do not contribute to \eqref{eq:ff34534}. Therefore we obtain
   \begin{align}
    & \frac{1}{2|eB|}\lim_{W\to 0}\frac{d}{dW} \int \frac{d^2q_\perp}{(2\pi)^2} \tr [\tilde G_{p+q}\gamma_5 \tilde G_q \gamma_5 ]
    \\ 
    & = \frac{F_0(q_4,q_3)}{2\pi}\Big( 
    \underbrace{g_1}_{m=n+1}+\underbrace{g_2}_{m=n-1}+\underbrace{g_3}_{m=n}
    \Big) + \frac{2|eB|}{\pi}\Big( 
    \underbrace{g_4}_{m=n+1}+\underbrace{g_5}_{m=n-1}+\underbrace{g_6}_{m=n} 
    \Big)\,,
    \label{eq:gg23}
   \end{align}
   where
   \begin{align}
    & g_1 = \sum_{n=0}^{\infty}
    \frac{ 2n+1 }{F_{n+1}(q_4,q_3)F_n(q_4,q_3)}
    \,, ~~
    g_2 = g_1\,, ~~
    g_3 = - \frac{1}{F_0(q_4,q_3)^2} - \sum_{n=1}^{\infty}
    \frac{4n}{F_n(q_4,q_3)^2}\,,
    \\
    & g_4 = \sum_{n=1}^{\infty}
    \frac{ n(n+1) }{F_{n+1}(q_4,q_3)F_n(q_4,q_3)}\,, ~~
    g_5 = g_4\,,  ~~ 
    g_6 = - \sum_{n=1}^{\infty}   \frac{2n^2}{F_n(q_4,q_3)^2}\,.
   \end{align}
   Plugging all into \eqref{eq:gg23}, we get
   \begin{align}
    & \frac{1}{2|eB|}\lim_{W\to 0}\frac{d}{dW} \int \frac{d^2q_\perp}{(2\pi)^2} \tr [\tilde G_{p+q}\gamma_5 \tilde G_q \gamma_5 ]
    \notag
    \\
    & =  \frac{1}{2\pi}\left(
    - \frac{1}{F_0(q_4,q_3)} + \frac{2}{F_1(q_4,q_3)}\right) + 
    \frac{F_0(q_4,q_3)}{\pi}
    \sum_{n=1}^{\infty}\left(
    \frac{ 2n+1 }{F_{n+1}(q_4,q_3)F_n(q_4,q_3)} 
    - \frac{2n}{F_n(q_4,q_3)^2}\right) 
    \notag
    \\
    & \quad + \frac{4|eB|}{\pi}
    \sum_{n=1}^{\infty}\left(
    \frac{ n(n+1) }{F_{n+1}(q_4,q_3)F_n(q_4,q_3)}
    - \frac{n^2}{F_n(q_4,q_3)^2}
    \right)
    \intertext{\noindent 
    where we have deliberately grouped the series into parentheses so that 
    the sums are convergent. We performed these sums over $n$ 
    with \emph{Mathematica}, finding
    }
    & = \frac{1}{2\pi}\left(
    - \frac{1}{F_0(q_4,q_3)} + \frac{2}{F_1(q_4,q_3)}\right) + 
    \frac{F_0(q_4,q_3)}{\pi}
    \frac{1}{4|eB|^2}\left\{
    2D\psi^{(1)}(1+D) + \frac{1}{1+D} - 2 
    \right\}
    \notag
    \\
    & \quad  +  \frac{4|eB|}{\pi}
    \frac{1}{4|eB|^2}
    \left\{ D-D^2\psi^{(1)}(1+D) \right\}
    \qquad \qquad  \left(\displaystyle D\equiv \frac{F_0(q_4,q_3)}{2|eB|} \right)
    \\
    & =  - \frac{1}{2\pi F_0(q_4,q_3)} + \frac{1}{2\pi |eB|} \,, 
   \end{align}
   where $\psi^{(1)}(x)$ is the first derivative of the digamma function.  

   Using $\tilde \del_k F_0(q_4,q_3) = 2k\, \theta(k^2 - q_3^2)$ one can  
   easily show
   \begin{align}
    \tilde \del_k\left( \frac{1}{2|eB|} \lim_{W\to 0}\frac{d}{d W} 
    \int \frac{d^2q_\perp}{(2\pi)^2} 
    \tr [\tilde G_{p+q}\gamma_5 \tilde G_q \gamma_5 ] \right) 
    = 
    \frac{1}{\pi}\frac{k}{(q_4^2+k^2+2g^2\bar\rho_k)^2}\, \theta(k^2 - q_3^2)\,.
   \end{align}
   Plugging this into \eqref{eq:ff34534} we finally obtain
   \begin{align}
    \del_k Z_k^\perp \Big|_{\rm fermi} & = - N_c g^2 \, 
    T\!\! \sum_{q_4:\,\text{odd}} \int \frac{dq_3}{2\pi} 
    \left[ \frac{1}{\pi}\frac{k}{(q_4^2+k^2+2g^2\bar\rho_k)^2}\, \theta(k^2 - q_3^2) \right]
    \\
    & = - \frac{1}{\pi^2} N_cg^2 k^2\, T\!\!\sum_{q_4:\,\text{odd}}\frac{1}{(q_4^2+k^2+2g^2\bar\rho_k)^2}
    \label{eq:z3432}
    \\
    & = 
    - \frac{1}{\pi^2} N_c g^2 k^2 \left(  
    \frac{1}{4 E_0(\bar\rho_k)^3}\tanh \frac{E_0(\bar\rho_k)}{2T} - 
    \frac{1}{8 E_0(\bar\rho_k)^2 T}
    \mathrm{sech}^2 \frac{E_0(\bar\rho_k)}{2T} 
    \right) \,,
   \end{align}
   with $E_0(\bar\rho_k) = \sqrt{k^2+2g^2\bar\rho_k\,}$\,.  The sum of
   \eqref{eq:z3432} and \eqref{eq:Zpp} yields $\del_k Z_k^\perp$ in
   \eqref{eq:fin_dZperp}.

   \subsubsection{Flow of $Z_k^\parallel$}

   Let us rewrite \eqref{eq:Zflow2} as
   \begin{align}
    \del_k Z_k^\parallel \Big|_{\rm fermi} 
    & = - \frac{1}{2} N_c g^2 \,T\!\! \sum_{q_4:\,\text{odd}} 
    \lim_{p\to 0}\frac{\del^2}{\del p_3^2} \int\frac{dq_3}{2\pi}
    \ \tilde \del_k \int \frac{d^2q_\perp}{(2\pi)^2}
    \tr [\tilde G_{p+q}\gamma_5 \tilde G_q \gamma_5 ]\,. 
    \label{eq:xxx_}
   \end{align}
   We shall carry out the calculations on the RHS in this order.  Let us
   take the limit $p_1\to 0$ (i.e., $W\to 0$) to focus on the
   $p_3$-dependence. Using \eqref{eq:Lagu} one can easily show that all
   terms in \eqref{eq:113-} with $|n-m|\geq 1$ vanish as $W\to 0$,
   leaving
   \begin{align}
    \int \frac{d^2q_\perp}{(2\pi)^2} \tr [\tilde G_{p+q}\gamma_5 \tilde G_q \gamma_5 ] \Big|_{W\to 0}
    = \sum_{n=0}^{\infty}
    \frac{1}{F_n(q_4,\qq_3)F_n(q_4,q_3)} 
    \left\{
    \alpha_n\frac{|eB|}{\pi} \FF_0(q_4,q_3,\qq_3) + \frac{4}{\pi} |eB|^2 n
    \right\}
    .\label{eq:105+-}
   \end{align}
   Then
   \begin{align}
    \int\frac{dq_3}{2\pi}\ \tilde \del_k 
    \left( 
    \int \frac{d^2q_\perp}{(2\pi)^2} \tr [\tilde G_{p+q}\gamma_5 \tilde G_q \gamma_5 ]  \Big|_{W\to 0} \right)
    & = \mathfrak{X}_1 + \mathfrak{X}_2 + \mathfrak{X}_3\,,
   \end{align}
   with the definitions 
   \begin{align}
    \mathfrak{X}_1 & \equiv - \int\frac{dq_3}{2\pi} 
    \sum_{n=0}^{\infty}
    \frac{\tilde\del_k F_n(q_4,\qq_3)}{F_n(q_4,\qq_3)^2F_n(q_4,q_3)} 
    \left\{
    \alpha_n \frac{|eB|}{\pi} \FF_0(q_4,q_3,\qq_3) + \frac{4}{\pi} |eB|^2 n
    \right\}\,,
    \\
    \mathfrak{X}_2 & \equiv - \int\frac{dq_3}{2\pi} 
    \sum_{n=0}^{\infty}
    \frac{\tilde\del_k F_n(q_4,q_3)}{F_n(q_4,\qq_3)F_n(q_4,q_3)^2} 
    \left\{
    \alpha_n \frac{|eB|}{\pi} \FF_0(q_4,q_3,\qq_3) + \frac{4}{\pi} |eB|^2 n
    \right\}\,,
    \\
    \mathfrak{X}_3 & \equiv \int\frac{dq_3}{2\pi} 
    \sum_{n=0}^{\infty}
    \frac{1}{F_n(q_4,\qq_3)F_n(q_4,q_3)} 
    \alpha_n \frac{|eB|}{\pi} \tilde\del_k\FF_0(q_4,q_3,\qq_3) \,. 
   \end{align}
   After elementary but quite lengthy calculations, we obtain (assuming
   $p_3>0$)
   \begin{align}
    \mathfrak{X}_1 & = 
    \scalebox{0.95}{$\displaystyle 
    - \frac{1}{\pi^2} |eB| k (2k-p_3)  
    \sum_{n=0}^{\infty}
    \frac{
    \alpha_n (q_4^2+k^2+2g^2\bar\rho_k)  + 4 |eB| n
    }{[q_4^2+E_n(\bar\rho_k)^2]^3} 
    + \frac{2}{\pi^2}|eB| k^3 p_3 \sum_{n=0}^{\infty}
    \frac{\alpha_n}{[q_4^2+E_n(\bar\rho_k)^2]^3}  
    $}
    \notag
    \\
    & \quad 
    - \frac{2}{\pi}|eB|k \int_{-k-p_3}^{-k}\frac{dq_3}{2\pi} 
    \sum_{n=0}^{\infty}
    \frac{ \alpha_n (q_4^2-kq_3+2g^2\bar\rho_k) + 4 |eB| n }
    {[q_4^2+E_n(\bar\rho_k)^2]^2 (q_4^2+q_3^2+2|eB|n+2g^2\bar\rho_k)} 
    \,,
    \label{eq:X_1}
    \\
    \mathfrak{X}_2 & = \mathfrak{X}_1\,,
    \\
    \mathfrak{X}_3 & = 
    - \frac{2}{\pi}|eB| \int_{-k-p_3}^{-k}\frac{dq_3}{2\pi} 
    \sum_{n=0}^{\infty}
    \frac{\alpha_n q_3}{[q_4^2+E_n(\bar\rho_k)^2]
    (q_4^2+q_3^2+2|eB|n+2g^2\bar\rho_k)} 
    \notag
    \\
    & \quad + 
    \frac{1}{\pi^2} |eB| k(2k-3p_3)  
    \sum_{n=0}^{\infty}\frac{\alpha_n }{[q_4^2+E_n(\bar\rho_k)^2]^2} 
    \,,  \label{eq:X_3}
   \end{align}
   where we have used $E_n(\rho)$ defined in \eqref{eq:alp-E}, and used 
   \begin{align}
    \tilde \del_k F_n(q_4,q_3) 
    & = 2k\, \theta(k^2 - q_3^2)\,, 
    \\
    \tilde \del_k F_n(q_4,\qq_3) & = 2k\, \theta(k^2 - \qq_3^2)\,,
    \\
    \tilde \del_k \FF_0(q_4,q_3,\qq_3)  
    &= \qq_3 q_3 \left\{
    \frac{\theta(k^2-q_3^2)}{|q_3|}[1+r_k(\qq_3)] + \frac{\theta(k^2-\qq_3^2)}{|\qq_3|}[1+r_k(q_3)]
    \right\}\,. 
   \end{align} 
   Note that the first line of \eqref{eq:X_1} and the second line of
   \eqref{eq:X_3} vanish in the limit $\lim_{p\to 0}\frac{\del^2}{\del
   p_3^2}$.  Using this fact, we find
   \begin{align}
    \lim_{p\to 0}\frac{\del^2}{\del p_3^2} 
    (\mathfrak{X}_1 + \mathfrak{X}_2 + \mathfrak{X}_3) 
    = 
    \frac{1}{\pi^2}|eB| \sum_{n=0}^{\infty} 
    \frac{\alpha_n}{[q_4^2+E_n(\bar\rho_k)^2]^2}\,.
   \end{align}
   Substituting this into \eqref{eq:xxx_} we finally arrive at
   \begin{align}
    \del_k Z_k^\parallel \Big|_{\rm fermi}  & = 
    -\frac{1}{2\pi^2} N_c g^2 |eB|\,T \!\!\sum_{q_4:\,\text{odd}}  \ 
    \sum_{n=0}^{\infty} \frac{\alpha_n}{[q_4^2 + E_n(\bar\rho_k)^2]^2} \,.
    \label{eq:dz8947}
   \end{align}
   The sum of \eqref{eq:dz8947} and \eqref{eq:dZ_b2} yields $\del_k
   Z_k^\parallel$ in \eqref{eq:fin_dZpara}.

   To speed up numerical computation we analytically summed over $n$,
   with the result
   \begin{multline}
    \del_k Z_k^\parallel \Big|_{\rm fermi} = 
    -\frac{N_c}{2\pi^2}g^2|eB|\Bigg\{  
    \frac{1}{4 E_0(\bar\rho_k)^3}\tanh \frac{E_0(\bar\rho_k)}{2T} - 
    \frac{1}{8 E_0(\bar\rho_k)^2 T} \mathrm{sech}^2 \frac{E_0(\bar\rho_k)}{2T} 
    \\
    + \frac{T}{2|eB|^2} \sum_{q_4:\,\text{odd}} \psi^{(1)} \left(1+\frac{q_4^2 + E_0(\bar\rho_k)^2}{2|eB|}\right)
    \Bigg\}\,. 
   \end{multline} 
   Since $\psi^{(1)}(x) \sim 1/x$ for $x\gg 1$, the sum is
   convergent.

 \section{Flow of the Taylor coefficients of $U_k$}
 \label{sc:taylorU}
 
 The flows \eqref{eq:taylor_flow} of parameters in the Taylor expansion
 of $U_k$ depend on $\del_k U'_k\Big|_{\bar\rho_k}$ and $\del_k
 U''_k\Big|_{\bar\rho_k}$.  The latter can be obtained from
 \eqref{eq:fin_dU} by taking the derivative with $\rho$ and substituting
 the polynomial expression \eqref{eq:U'_}.  After elementary
 calculations, we arrive at
 \begin{align} 
  & \del_k U'_k\Big|_{\bar\rho_k} 
  \notag
  \\
  & = 
  - \frac{k^2}{8\pi^2}\left(1+\frac{k}{3}\frac{\del_k Z_k^\parallel}{Z_k^\parallel}\right)
  \frac{a_k^{(2)}}{Z_k^\perp}\left\{
  \frac{\coth\left(\frac{1}{2T}\sqrt{k^2+\frac{a_k^{(1)}}{Z_k^\parallel}}\right)}{\sqrt{k^2+\frac{a_k^{(1)}}{Z_k^\parallel}}}
  + 3 \,
  \frac{\coth\left(\frac{1}{2T}\sqrt{k^2+\frac{a_k^{(1)}+2\bar\rho_k
  a_k^{(2)}}{Z_k^\parallel}}\right)}{\sqrt{k^2+\frac{a_k^{(1)}+2\bar\rho_k
  a_k^{(2)}}{Z_k^\parallel}}}
  \right\} 
  \notag
  \\
  & \quad 
  - \frac{N_c}{2\pi^2}g^2k^2|eB| \sum_{n=0}^{\infty}\alpha_n
  \left( \frac{\mathrm{sech}^2 \frac{E_n(\bar\rho_k)}{2T}}{2T\cdot E_n(\bar\rho_k)^2} - 
  \frac{\tanh \frac{E_n(\bar\rho_k)}{2T}}{E_n(\bar\rho_k)^3} \right)   \,,
  \label{eq:U_1_pol}
 \end{align}
 and
 \begin{align} 
  & \del_k U''_k\Big|_{\bar\rho_k} 
  \notag
  \\
  & = \frac{k^2}{8\pi^2}\left(1+\frac{k}{3}\frac{\del_k
  Z_k^\parallel}{Z_k^\parallel}\right)
  \frac{\big(a_k^{(2)}\big)^2}{2 Z_k^\perp Z_k^\parallel }
  \left\{
  \frac{\coth \left(\frac{1}{2T}\sqrt{
  k^2+\frac{a_k^{(1)}}{Z_k^\parallel}  }\right) }{ \left(
  k^2+\frac{a_k^{(1)}}{Z_k^\parallel} \right)^{3/2}}
  +
  \frac{\mathrm{csch}^2\left(\frac{1}{2T}\sqrt{k^2+\frac{a_k^{(1)}}{Z_k^\parallel}}\right)}{2T
  \left(k^2+\frac{a_k^{(1)}}{Z_k^\parallel}\right)}
  \right.
  \notag
  \\
  & \qquad \qquad \left. + ~
  \frac{9 \coth
  \left(\frac{1}{2T}\sqrt{k^2+\frac{a_k^{(1)}+2\bar\rho_ka_k^{(2)}}{Z_k^\parallel}}\right)}{
  \left(k^2+\frac{a_k^{(1)}+2\bar\rho_ka_k^{(2)}}{Z_k^\parallel}\right)^{3/2}}
  + \frac{9~\mathrm{csch}^2
  \left(\frac{1}{2T}\sqrt{k^2+\frac{a_k^{(1)}+2\bar\rho_ka_k^{(2)}}{Z_k^\parallel}}
  \right)}{2T\left(k^2+\frac{a_k^{(1)}+2\bar\rho_ka_k^{(2)}}{Z_k^\parallel}\right)}
  \right\}
  \notag
  \\
  & \quad  - \frac{N_c}{2\pi^2}g^4k^2|eB| \sum_{n=0}^\infty \alpha_n 
  \left( 
  -\frac{3}{2T}\frac{\mathrm{sech}^2\frac{E_n(\bar\rho_k)}{2T}}{E_n(\bar\rho_k)^4}
  + 3 \frac{\tanh\frac{E_n(\bar\rho_k)}{2T}}{E_n(\bar\rho_k)^5} 
  - \frac{1}{2T^2}\frac{\mathrm{sech}^2\frac{E_n(\bar\rho_k)}{2T}\tanh \frac{E_n(\bar\rho_k)}{2T}}{E_n(\bar\rho_k)^3}
  \right)   
  \,.  \label{eq:U_2_pol}
 \end{align}
 The convergence of the Landau level sums in \eqref{eq:U_1_pol} and
 \eqref{eq:U_2_pol} is rather slow, due to the terms $\frac{\tanh
 \frac{E_n(\bar\rho_k)}{2T}}{E_n(\bar\rho_k)^3}$ and
 $\frac{\tanh\frac{E_n(\bar\rho_k)}{2T}}{E_n(\bar\rho_k)^5}$ that decay
 only slowly especially when $k^2+2g^2\bar\rho_k \gg 2|eB|$.  From a
 computational point of view, it is advantageous to split the zero
 temperature part from the thermal part as
 $\tanh\frac{E_n(\bar\rho_k)}{2T} = 1 +
 \Big(\tanh\frac{E_n(\bar\rho_k)}{2T}-1\Big)$ and perform the summation
 in the zero temperature part analytically as 
 \begin{align}
  \sum_{n=0}^{\infty} \frac{\alpha_n}{E_n(\bar\rho_k)^3} & = 
  - \frac{1}{E_0(\bar\rho_k)^3} +
  \frac{1}{\sqrt{2}|eB|^{3/2}}\zeta\left( \frac{3}{2},
  \frac{E_0(\bar\rho_k)^2}{2|eB|} \right)\,,
  \\
  \sum_{n=0}^{\infty} \frac{\alpha_n}{E_n(\bar\rho_k)^5} & = 
  - \frac{1}{E_0(\bar\rho_k)^5} +
  \frac{1}{2\sqrt{2}|eB|^{5/2}}\zeta\left( \frac{5}{2},
  \frac{E_0(\bar\rho_k)^2}{2|eB|} \right)\,,
 \end{align}
 where $\zeta(x,y)$ is the Hurwitz zeta function. Then all the terms in
 the remainder are suppressed by a Boltzmann factor $\sim
 \ee^{-E_n(\bar\rho_k)/T}$ and only a small number of Landau levels
 contribute to the sum. We found that this trick speeds up numerical
 computation of the flow equation considerably.

 \bibliography{draft_frg_v2.bbl}
\end{document}